\documentclass[showpacs, aps, prb, unsortedaddress,showkeys,longbibliography]{revtex4-1}

\usepackage{amssymb}
\usepackage{latexsym}
\usepackage{amsmath}
\usepackage{graphicx}
\usepackage{float}
\usepackage{xcolor}

\begin{document}

\title{Scattering under Linear Non Self-Adjoint Operators: Case of in-Plane Elastic Waves}

\date{\today}
\author{Amir Ashkan Mokhtari$^1$}
\author{Yan Lu$^2$}
\author{Qiyuan Zhou$^1$}
\author{Alireza V. Amirkhizi$^3$}
\author{Ankit Srivastava$^1$}
\thanks{Corresponding Author}  
\affiliation{$^1$ Department of Mechanical, Materials, and Aerospace Engineering,
Illinois Institute of Technology, Chicago, IL, 60616
USA}
\email{asriva13@iit.edu}
\affiliation{$^2$ Department of Mechanical Engineering, Boston University}
\affiliation{$^3$ Department of Mechanical Engineering, University of Massachusetts Lowell}

\begin{abstract}
In this paper, we consider the problem of the scattering of in-plane waves at an interface between a homogeneous medium and a metamaterial. The relevant eigenmodes in the two regions are calculated by solving a recently described non self-adjoint eigenvalue problem particularly suited to scattering studies. The method efficiently produces all propagating and evanescent modes consistent with the application of Snell's law and is applicable to very general scattering problems. In a model composite, we elucidate the emergence of a rich spectrum of eigenvalue degeneracies. These degeneracies appear in both the complex and real domains of the wave-vector. However, since this problem is non self-adjoint, these degeneracies generally represent a coalescing of both the eigenvalues and eigenvectors (exceptional points). Through explicit calculations of Poynting vector, we point out an intriguing phenomenon: there always appears to be an abrupt change in the sign of the refraction angle of the wave on two sides of an exceptional point. Furthermore, the presence of these degeneracies, in some cases, hints at fast changes in the scattered field as the incident angle is changed by small amounts. We calculate these scattered fields through a novel application of the Betti-Rayleigh reciprocity theorem. We present several numerical examples showing a rich scattering spectrum. In one particularly intriguing example, we point out wave behavior which may be related to the phenomenon of resonance trapping. We also show that there exists a deep connection between energy flux conservation and the biorthogonality relationship of the non self-adjoint problem. The proof applies to the general class of scattering problems involving elastic waves (under self-adjoint or non self-adjoint operators).
\end{abstract}

\maketitle

\section{Introduction}
There has been considerable recent research interest in the field of wave propagation through periodic structures under the fields of photonics, phononics, and metamaterials\cite{nemat2015anti,shmuel2016universality,chen2016modulating,MOKHTARI2019256, Amirkhizi2018, Hajarolasvadi2019, Nejadsadeghi2019,Misra2019,huang2009negative,Huang2014,Liu2011}. Irrespective of the different properties which metamaterials research in different fields target, the end goal is the same: to design composite materials for the fine-tuned, predominantly frequency-dependent control of the trajectory and dissipation characteristics of waves\cite{hussein2014dynamics, hussein2007dispersive, norris2008acoustic}.

A periodically layered composite is an example of a simple but physically rich phononic system for which there is already considerable research in the area of dispersion properties \cite{nayfeh1995wave,willis2015negative,NematNasser2015,norris1993waves, norris1992dispersive,srivastava2016metamaterial,MOKHTARI2019256,amirkhizi2017homogenization,Morini2019,Li2018,Li2019}. This problem can be made more complicated by interfacing a semi-infinite layered composite with a homogeneous material. This problem has been studied by several researchers recently for the anti-plane shear case \cite{willis2015negative, srivastava2017evanescent, srivastava2016metamaterial,sharma2019wave}. Even for this simple problem, exotic wave behaviors such as negative refraction and negative effective properties have been reported \cite{willis2015negative,NematNasser2015,srivastava2016metamaterial}. The problem can be further complicated by considering in-plane waves instead of anti-plane shear waves. Mokhtari et. al.\cite{mokhtari2019properties} showed that the associated eigenvalue problem in this case is non-normal which gives rise to complex eigenvalues (in general) and non orthogonal eigenvectors. It has been shown in earlier studies that the non-orthogonality of the eigenvectors in such problems is associated with interesting physical phenomena \cite{Makris2008,Kostenbauder1997}. The scattering problem involving the layered system under consideration here was recently considered by Lustig et. al.\cite{Gal2019}. They made some important observations in their paper, perhaps the most important being the explicit identification of eigenvalue degeneracies (exceptional points, EPs) in the dispersion spectrum of the composite. EPs are degeneracies in the spectrum of non self-adjoint operators, \cite{lu2018level,heiss1990avoided,heiss2000repulsion} and are associated with anomalous scattering behavior \cite{lu2018level,Ji2019}. They also noted the important point that in this problem, the exceptional points are being generated without any gain or loss in the system (unlike PT symmetric studies\cite{ElGanainy2018,Zhu2014}) and that they are accessible through an appropriate scattering problem. They solved a limited scattering problem, however, it was not in the vicinity of the exceptional points that they identified.

In this paper, we consider the same problem that Lustig et. al.\cite{Gal2019} considered but make several novel contributions and observations. We exploit the linear non self-adjoint eigenvalue problem given in Mokhtari et. al. \cite{mokhtari2019properties} directly to evaluate the complex dispersion relation of the composite. It gives us the ability to characterize the EPs present in the spectrum more efficiently than the Transfer Matrix Method \cite{willis2015negative,srivastava2017evanescent} which has been the method of choice in previous related studies in layered systems. We identify a new class of EPs in the problem (in the domain of the wavevector component parallel to the layers) which were not identified previously. We identify these EPs at the intersections of  propagating and non-propagating branches (Figs. \ref{k1EP}a and \ref{fig-EPk2}a), and also at intersections of two propagating branches (Fig. \ref{realEP}a). Through explicit numerical calculations of the Poynting vector, we show an intriguing property of wave propagation: there always appears to be an abrupt change in the sign of refraction angle of the wave on two sides of an EP. This seems to hold for EPs both in the domain of the wavenumber parallel to the layers and perpendicular to the layers (Fig. \ref{refraction}). 

Solving the scattering problem requires the calculation of the scattering coefficients. The classic way \cite{achenbach1984wave,willis2015negative,srivastava2017evanescent,2002.00739} of solving this problem in the anti-plane shear case is through an appropriate discretization of the continuity conditions by employing the orthogonality of modeshapes. However, the modeshapes in the in-plane problem are non-orthogonal which causes the solution process to become numerically unstable. We solve this problem here through a novel application of the Betti-Rayleigh reciprocity theorem \cite{Achenbach2004}. We show scattering calculations for several cases, elucidating a rich spectrum of scattering behavior. This includes showing fast variations in the scattering spectrum in the vicinity of EPs and, in one intriguing case, a sudden drop in transmission energy in what may be a sign of the resonance trapping phenomenon \cite{Mller2009,Persson2000}. Finally, we give an important proof regarding energy flux conservation in scattering problems under non-normal operators: we show that the energy flux component applicable to conservation is a restatement of the biorthogonality relationship of the problem. This observation allows us to diagonalize the flux conservation relationship, elucidating the interplay between the scattering coefficients of the left and right eigenvectors. We note the important result that such diagonalization is impossible without the use of biorthogonality. 

This paper is organized as follows: Section II describes the engineering motivations of solving the scattering problem and important practical applications related to EPs. Section III presents the problem statement and formulation. Section IV discusses the dispersion calculations using TMM and linear non self-adjoint formulation in addition to the existence of exceptional points. Section V presents the application of Betti-Rayleigh reciprocity in solving the scattering problem, and also the proof of direct connection between biorthoginality and energy flux conservation. Section VI presents numerical examples, including scattering calculations supported by FEM simulations.

\section{Engineering motivation}
\begin{figure}[htp]
\centering
\includegraphics[scale=.4]{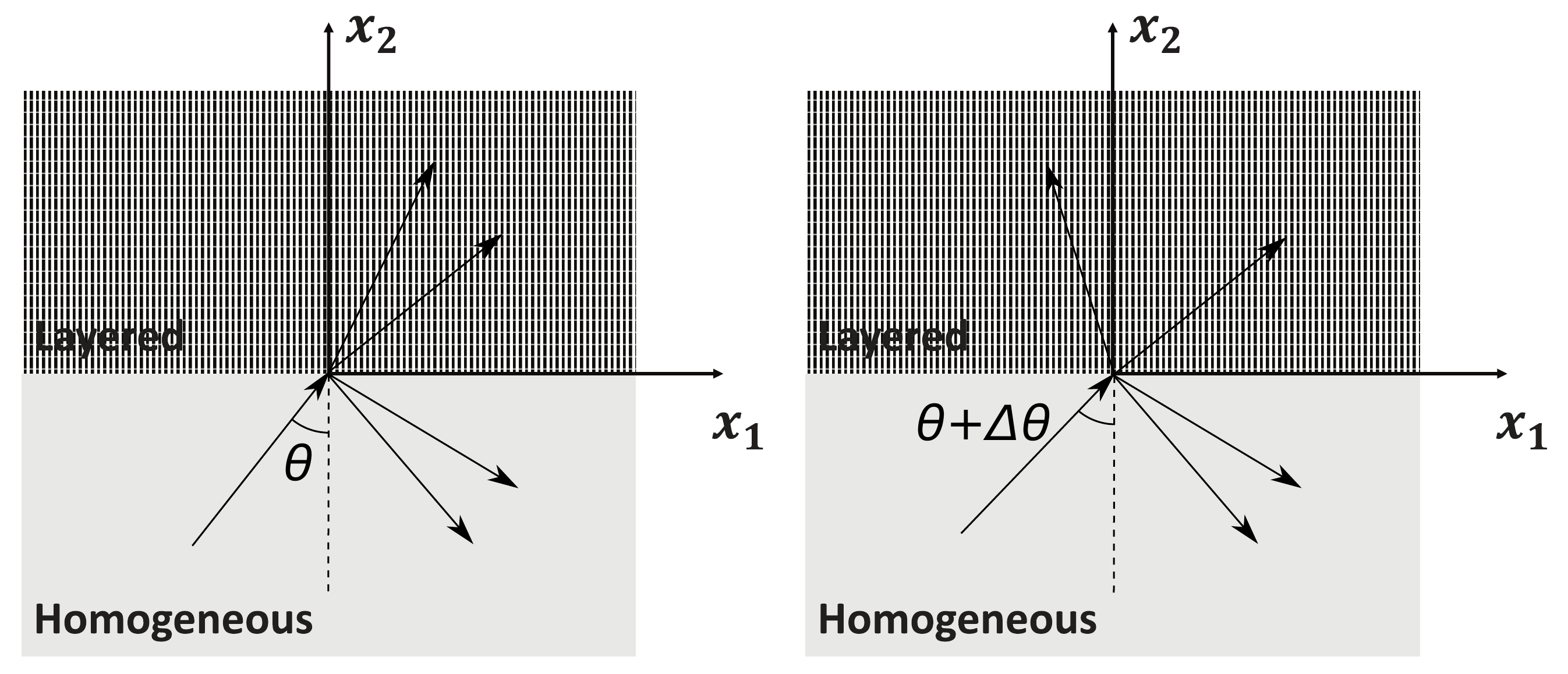}
\caption{Schematic of the single interface problem showing an incident wave and some reflected and transmitted waves. The exact number of reflected and transmitted waves depends upon the problem.}\label{fschematic}
\end{figure}
Non self-adjoint operators usually describe systems that exchange energy with their environment through interfaces. Such systems have been extensively studied recently in the areas of wave propagation\cite{achilleos2017non}, vibration \cite{udwadia2012longitudinal} and quantum systems\cite{gao2015observation}. Two recent studies in phononics elucidate the interplay between self-adjoint systems and non self-adjoint systems. Lu et. al. \cite{lu2018level} showed that non self-adjoint systems emerge when one simply extends the wavenumber dependent equilibrium equations into the complex parameter domain and that it leads to interesting physical and related \cite{heiss2000repulsion} effects of level repulsion and exceptional points. Mokhtari et. al. \cite{mokhtari2019properties} showed how to formulate general phononic eigenvalue problems in self-adjoint and non self-adjoint forms. These conversions offers access to the exotic dynamic behavior around level repulsion \cite{frank1994classical,amirkhizi2018reduced} and exceptional points \cite{miri2019exceptional}.

Level repulsion, sometimes termed avoided crossing\cite{heiss1990avoided} or mode veering\cite{manconi2017veering} in different research communities, appears in a parameter dependent eigensystems when there exists mode coupling between two eigenstates. As the mode coupling strength increases, the two eigenfrequency branches involved veer away from each other and the corresponding modeshapes undergo a swift exchange under a small variation of the parameter. This effect can be used to create highly sensitive sensors which measure the perturbation of the parameter upon which the fast modeshape changes depend. In the past, researchers have considered perturbations on stiffness \cite{manav2014ultrasensitive}, geometric factors\cite{foreman2013level,dehrouyeh2016free}, pressure \cite{alcheikh2019highly}, and electrothermal voltage \cite{hajjaj2017static} and hypothesised sensors based on these perturbations.

Exceptional points, on the other hand, are characterized by non-linear dispersion in their vicinity \cite{hodaei2017enhanced}, modeshape sign flips \cite{heiss1999phases}, and non-trivial geometric phase accumulation \cite{mailybaev2005geometric}. These unusual properties can be used for the design of sensors as well. The design principles for exceptional point based sensors can be divided into two categories, one that relies on the measurement of nonlinear dispersion, and the other that relies on the sensitivity of modeshape. In the former category, Alcheikh et. al. designed a pressure sensor based on tracking the resonant frequency of a micro-beam under varying axial load in the vicinity of an exceptional point \cite{alcheikh2019highly}. Similarly, by introducing higher order exceptional points into a photonic system, Hodaei et. al. \cite{hodaei2017enhanced} were able to enhance the sensitivity of the measurements of resonant frequencies of the system. In the latter category, a number of MEMS sensor designs exploit mode localization behavior where a symmetry breaking perturbation leads to vibration energy confinement \cite{thiruvenkatanathan2009effects,manav2014ultrasensitive,zhang2017algebraic,zhang2018suppression,zhang2018linear,rabenimanana2019mass}. In these cases, symmetry breaking appears in level repulsion zones in the vicinity of exceptional points and modeshape sensitivity becomes orders of magnitude larger compared to resonant frequency sensitivity. In addition to the design of level repulsion/exceptional points based sensors in the classic mechanics settings, the physics of such phenomenon has been shown to have promise in enhancing quantum sensing as well \cite{lang2015dynamical,samutpraphoot2020strong}.

As in other areas, it is possible to create highly sensitive sensors for wave propagation applications using the physics of level repulsion and exceptional points. Fig. (\ref{fschematic}) shows a schematic of how one such sensor might work. It shows an interface between a homogeneous medium and a layered medium (phononic crystal). An incident wave at an angle $\theta$ (with a wavenumber $k$ and frequency $\omega$) produces potentially multiple reflected and transmitted modes. Snell's law ensures that the tangential component of the wavevector, $k_1=k\sin\theta$, is conserved for all the reflected and transmitted modes. In this problem, the modes which exist in the transmitted and reflected sides depend upon the common parameter $k_1$ which can be changed by changing the angle of incidence of the incident wave. Since the angle of incidence depends upon the location of the source which is producing the incident wave, a sensitive measurement of $k_1$ is equivalent to a sensitive localization of the source. The geometry of the phononic crystal may be designed to create exceptional points and level repulsion zones in the parameter of $k_1$ thus giving us the potential ability of measuring $k_1$ with high sensitivity. This is also schematically shown in Fig. (\ref{fschematic}). The idea is to design the phononic crystal in such a way that when the incident angle is changed by a small amount, it leads to a large change in the vibratory modes of the crystal thus altering the scattered field significantly. A specific possibility is one where a small increment in the source incident angle transforms one transmitted mode from positive refraction to negative refraction (Fig. \ref{fschematic}). As will be shown later in the numerical example section, this mode splitting occurs exactly around an exceptional point. In this design scheme, the sensing metric will be based on the scattered waves field of this problem. It is, therefore, important to understand this scattering problem, particularly under the non self-adjoint formalism as this problem is indeed non self-adjoint \cite{mokhtari2019properties}. In the future it will be of great interest to compare the performance of sensors based on the principles of level repulsion and exceptional points with traditional source localization method based on clusters of sensors and data processing algorithms\cite{kundu2014acoustic}. 

\section{Statement of the in-plane problem}
Following \cite{willis2015negative,srivastava2016metamaterial,srivastava2017evanescent,Gal2019}, we define our laminate as a periodically layered structure in the $x_1$ direction with the layer interfaces in the $x_2-x_3$ plane extending to infinity. In the direction of periodicity, the laminated composite is characterized by a unit cell $\Omega$ of length $h$ ($0\leq x_1\leq h$). For our purposes the unit cell is composed of 2 material layers with Lame's parameters $\lambda_1,\lambda_2,\mu_1,\mu_2$, density $\rho_1,\rho_2$, and thicknesses $h_1,h_2$ respectively. However, this restriction is not necessary. In fact, relaxing this may give rise to other interesting physical phenomena \cite{Gal2019}. The case of $n$ homogeneous material layers or layers with spatially changing material properties is not substantively different. All that is required is that the material properties be periodic with the unit cell so as to give the composite its phononic character.

We assume the problem to be in-plane in nature. If in-plane waves are propagating in the laminated or homogeneous medium, then the nonzero components of displacement are taken to be $u_1,u_2$. These have the functional form $u_1(x_1,x_2,t),u_2(x_1,x_2,t)$ and give rise to the relevant stress components $\sigma_{11},\sigma_{22},\sigma_{12}=\sigma_{21}$ with the same functional form as the displacements. By denoting any of these fields with the symbol $\Phi$, the periodicity of the problem enforces that the fields can be expressed in the following Bloch-periodic form:
\begin{eqnarray}
\label{eBloch}
\displaystyle \Phi(x_1,x_2,t)=\bar{\Phi}(x_1)\exp\left[i(k_1x_1+k_2x_2-\omega t)\right]
\end{eqnarray}
where $\bar{\Phi}$ is the unit cell periodic part of $\Phi$.

Here we are concerned with a scattering problem which involves an interface between a homogeneous medium with material properties $\lambda_0$, $\mu_0$ and a layered composite. The interface is infinite in the $x_1-x_3$ plane and located at $x_2=0$ (Fig. (\ref{fschematic})), The layered medium is in the region $x_2>0$ with the layers being parallel to the $x_2-x_3$ plane. A plane harmonic wave with frequency $\omega$ and $x_1$ wavenumber component $k_1$ is incident at the interface from the homogeneous medium and is of the form:
\begin{eqnarray}
\label{eIncident}
\displaystyle \mathbf{u}^I(x_1,x_2,t)=A\mathbf{d}\exp\left[i(k_1x_1+k_2x_2-\omega t)\right]
\end{eqnarray}
where the wavenumber is assumed to satisfy the dispersion relation of the homogeneous medium, $k_1,k_2\geq 0$, and $\mathbf{d}=[\sin(\theta)\ \cos(\theta)]$ for a longitudinal incident wave and $\mathbf{d}=[-\cos(\theta)\ \sin(\theta)]$ for a shear incident wave, where $\theta$ is the angle of incidence. This wave sets up an infinite number of transmitted and reflected waves. All these scattered waves share the same frequency and $x_1$ component of the wavevector as the incident wave. The transmitted and reflected fields have the following forms for the displacement vector:
\begin{eqnarray}
\label{eScattered}
\nonumber \displaystyle \mathbf{u}^T(x_1,x_2,t)=\sum_{i=1}^{\infty}T^{(i)}\bar{\mathbf{u}}_T^{(i)}(x_1)\exp\left[i(k_1x_1+k_2^{(i)} x_2-\omega t)\right]\\
\displaystyle \mathbf{u}^R(x_1,x_2,t)=\sum_{i=1}^{\infty}R^{(i)}\bar{\mathbf{u}}_R^{(i)}(x_1)\exp\left[i(k_1x_1+\kappa_2^{(i)} x_2-\omega t)\right]
\end{eqnarray}
where $T^{(i)},R^{(i)}$ are the scattering coefficients of the modes which are scattered in the transmitted and reflected regimes respectively. All these modes are characterized by the same $x_1$ wavenumber $k_1$ but different $x_2$ wavenumbers. The $x_2$ wavenumbers $k_2^{(i)},\kappa_2^{(i)}$ are chosen so as to ensure that either the flux of these waves is zero or away from the interface. This involves choosing those $k_2^{(i)}$ with $\Im{k_2^{(i)}}>0$ and $\kappa_2^{(i)}$ with $\Im{\kappa_2^{(i)}}<0$. If the imaginary parts are zero, then positive $k_2^{(i)}$ and negative $\kappa_2^{(i)}$ modes are chosen. $\bar{\mathbf{u}}_T^{(i)},\bar{\mathbf{u}}_R^{(i)}$ are the displacement modeshapes corresponding to the modes which are characterized by the $x_2$ wavenumbers $k_2^{(i)},\kappa_2^{(i)}$ respectively. 

\section{Dispersion Calculations}
The first step in determining scattering coefficients is to calculate the acceptable $x_2$ wavenumbers and modeshapes which correspond to a given frequency $\omega$ and a given $x_1$ wavenumber $k_1$. We will consider two methods of doing this: 
\begin{itemize}
    \item Transfer Matrix Method (TMM): This is an exact numerical method but it naturally only produces solutions of the form $k_1(\omega,k_2)$. This is the method that has been used in previous studies. \cite{willis2015negative,srivastava2016metamaterial} In this method, $k_2$ values for a desired $\omega,k_1$ combination must be determined through a search procedure. Furthermore, the corresponding modeshapes need to be separately calculated as well.
    \item Direct solution of a non self-adjoint eigenvalue problem \cite{mokhtari2019properties}: This method directly solves for $k_2(\omega,k_1)$ and automatically produces the corresponding modeshapes in both the homogeneous medium and the layered composite. In fact, this method can be used to directly calculate any $k_p(\omega,k_i)$ where $k_p,k_i$ are components along any two perpendicular directions.
\end{itemize}
Note that one could as well solve $\omega(k_1,k_2)$ problem but this is the common form in which phononic problems are generally considered and will not be considered in this paper. There is an important point of note in this discussion. It has been shown \cite{lu2018level,heiss1990avoided,heiss2000repulsion,Ji2019,Zhu2014} that the degeneracies of the phononic spectrum have important physical consequences for wave propagation. In the present problem, these degeneracies (exceptional points) can exist for either $\omega$, $k_1$, or $k_2$. Lustig et. al.\cite{Gal2019} has made the interesting point that in this simple elastic in-plane problem, it becomes possible to access these degeneracies in the wave-number domain. However, they have only pointed out the degeneracies for $k_1$ ($\omega$ degeneracies being unimportant in this case). We argue in the coming sections that it is the degeneracies in the $k_2$ spectrum which are more important (at least for the scattering problem considered in this paper and, to a large extent, even in Lustig et. al.\cite{Gal2019}) and which can be efficiently determined using our direct solution of the $k_2(\omega,k_1)$ problem. This is because if these degeneracies occur for a real $k_1$, they (and the complex $k_2$ modes close to the degeneracy) can be directly activated in the scattering problem.

\subsection{TMM and $k_1(\omega,k_2)$ solutions}
\begin{figure}[htp]
\centering
\includegraphics[scale=.55]{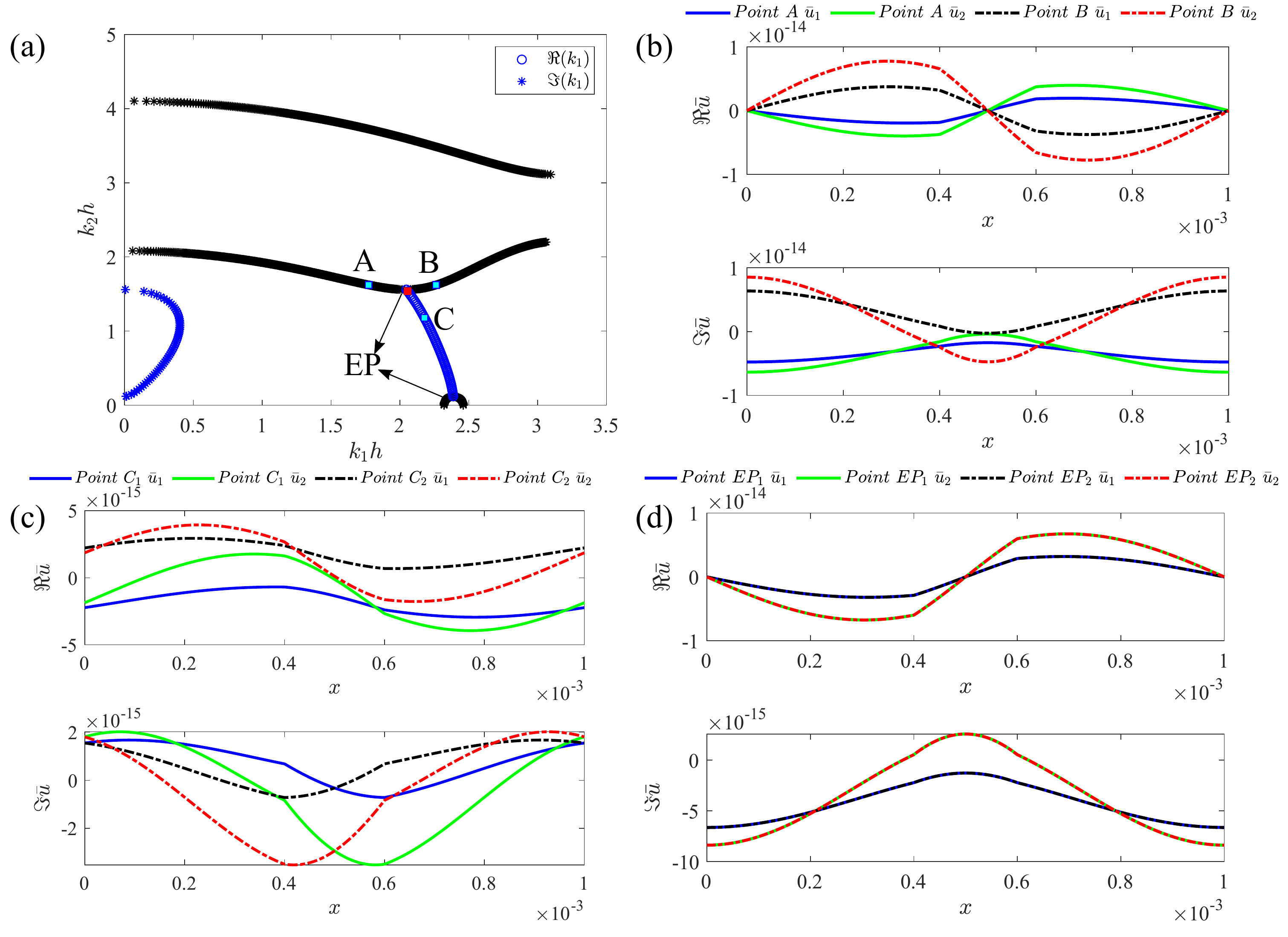}
\caption{(a) Exceptional points at $f=1.97$ MHz are marked by the two arrows. The EPs occur at the intersection of real branches and the branch with complex $k_1$ values (Blue curves), (b) Displacement and stress modeshapes at points A and B showing that the modeshapes are different, (c)
 Modeshapes at point C which represents two complex conjugate modes represented by $C_1,C_2$, (d) modeshapes at the EP at $k_2h=1.5588$ explicitly showing the coalescing of the modeshapes corresponding to the coalescing eigenvalues.}\label{k1EP}
\end{figure}

The transfer matrix method has been explained in detail in \cite{haque2017generalized} and it is treated only briefly here for completeness (see Appendix-A for some more details). Following \cite{haque2017generalized} we define our laminate as a periodically layered structure in the $x_1$ direction with the layer interfaces in the $x_2$--$x_3$ plane and infinite in this plane. In the direction of periodicity the laminated composite is characterized by a unit cell $\Omega$ of length $h$ ($0\leq x_1\leq h$). For our purposes the unit cell is composed of two material layers with Lam\'{e} constants $\lambda_1$, $\lambda_2$, $\mu_1$, $\mu_2$, densities $\rho_1$, $\rho_2$, and thicknesses $h_1$, $h_2$ respectively. 
The wave number solutions come from the following equation (Appendix-A)
\begin{equation}
\label{eInPlaneS}
\cos(k_1h)=\dfrac{1}{4}\left[a_3\pm\sqrt{a_3^2-4a_2+8}\right],
\end{equation}
where $a_0, a_1, a_2, a_3$ are defined in Eq. (\ref{app_param}). So that if $k_1$ is a solution, then so are $\pm(k_1\pm 2n\pi/h)$ for all integer $n$ (details in \cite{haque2017generalized}). This class of problem, termed $k_1(\omega,k_2)$, is generally associated with a non-normal differential operator \cite{mokhtari2019properties} This generally gives rise to complex eigenvalues and linearly dependent eigenvectors. TMM may be used to calculate exceptional points in the $k_1$ spectrum. For instance, consider the following composite:
\begin{itemize}
\item Material 1: $\lambda_1=121.1GPa$, $\mu_1=80.8GPa$, $\rho_1=7800kg/m^3$, $h_1=0.8mm$
\item Material 2: $\lambda_2=51.1GPa$, $\mu_2=26.3GPa$, $\rho_2=2700kg/m^3$, $h_2=0.2mm$
\end{itemize}
For this composite, the $k_1-k_2$ solutions at $f=1.97$ MHz are shown in Fig. (\ref{k1EP}a). In this figure, $k_2$ is considered real. Three branches of real $k_1$ modes are plotted in black, and a branch of complex $k_1$ modes connecting the lower two real modes is plotted in blue. Note that there are two fundamental $k_1$ solutions for each $k_2$ (Eq. (\ref{eInPlaneS})), therefore, the complex branch consists of complex conjugate pair solutions. Only the positive imaginary parts of the solutions are shown in Fig. (\ref{k1EP}a). The exceptional points occur at the two ends of the complex branch intersecting the real branches. At these points, the imaginary parts of the complex branch disappears. These exceptional points are characterized not only by a coalescing of the eigenvalues but also a corresponding coalescing of the eigenvectors. To see this explicitly, we plot the modeshapes for three points in the vicinity of an exceptional point (points $A,B,C$) as well as the modeshapes at the exceptional point itself. Points A and B are at $k_2h=1.6216$ and Point C is at $k_2h=1.1774$. Figs. (\ref{k1EP}b,\ref{k1EP}c) show that the modes in the immediate neighborhood of the exceptional point have distinct modeshapes. In contrast, the modeshapes at the EP at $k_2h=1.5588$ lie on top of each other, showing a coalescing of the eigenvectors. 

\subsection{Direct solution of the $k_2(\omega,k_1)$ problem}

Rather than solving the $k_1(\omega,k_2)$ problem as we do in the last section, it is possible to solve the $k_2(\omega,k_1)$ problem. However, such a solution is not possible through the Transfer Matrix Method. The strategy for such a solution is detailed in \cite{mokhtari2019properties} and here we present it only briefly. Consider the linear non self-adjoint form of a phononics eigenvalue problem:
\begin{eqnarray}
\label{Eq_motion}
\nabla . \boldsymbol{\sigma}=\rho \Ddot{\mathbf{u}}\\
\boldsymbol{\sigma}=\mathbf{C}:\nabla \mathbf{u}
\end{eqnarray}
where $\boldsymbol{\sigma}$ is the stress tensor and $\mathbf{u}=\{u_1, u_2\}$ is the displacement vector. Consider a kind of phononic eigenvalue problem in which we are given the frequency $\omega$ and one of the two components of the wavevector $\boldsymbol{k}=\{k_1,k_2\}$. The fields are of the form $\mathbf{u}=\bar{\mathbf{u}}\exp(i(k_ix_i-\omega t))$ and $\boldsymbol{\sigma}=\bar{\boldsymbol{\sigma}}\exp(i(k_ix_i-\omega t))$ where the barred quantities represent the unit cell periodic parts as explained earlier. Denoting by the roman indices $1,2$ we can separate the equations of motion into the following:
\begin{eqnarray}
\label{EStateSpace}
\displaystyle \nonumber \bar{\sigma}_{ij,j}-ik_1\bar{\sigma}_{i1}+\omega^2\rho \bar{u}_i=ik_2\bar{\sigma}_{i2}\\
 \bar{\sigma}_{ij}-C_{ijkl} \bar{u}_{k,l}+iC_{ijk1}k_1\bar{u}_{k}=-iC_{ijk2}k_2\bar{u}_{k}
 \label{mixed_form}
\end{eqnarray}
These equations can be cast in the generalized eigenvalue form by identifying two new vectors. Specifically, we consider $\boldsymbol{\gamma}=\{k_1, 0\}$ and $\mathbf{n}=k_2\{0,1\}$. This transforms Eq. (\ref{EStateSpace}) into the following form:
\begin{eqnarray}
\displaystyle \nonumber A\bar{\phi}=k_2 B\bar{\phi}
\end{eqnarray}
where $\bar{\phi}\equiv \{\bar{\mathbf{u}},\bar{\boldsymbol{\sigma}}\}$, and $A,B$ are defined as:
\begin{eqnarray}\label{Eq:beta3problem}
\displaystyle A=
\begin{bmatrix}
\omega^2\rho (\ ) & \boldsymbol{\nabla}\cdot (\ )-i(\ )\cdot\boldsymbol{\gamma}\\
-\mathbf{C}:\boldsymbol{\nabla}(\ )+i\mathbf{C}: (\ )\otimes\boldsymbol{\gamma} &\mathbf{I}
\end{bmatrix};\quad
\displaystyle B=
\begin{bmatrix}
0 & i(\ )\cdot \mathbf{n}\\
-i\mathbf{C}:(\ )\otimes\mathbf{n} & 0
\end{bmatrix}
\end{eqnarray}
The above non-self adjoint eigenvalue problem can be solved using the Plane Wave Expansion method whose details have been provided elsewhere \cite{mokhtari2019properties}. Importantly, the above eigenvalue problem directly provides us with a solution for $k_2$ given $(\omega,k_1)$. If desired, it is trivial to cast the above in a $k_1(\omega,k_2)$ form which would then provide the solutions calculated by the Transfer Matrix Method.

\subsection{$k_2$ exceptional points}

\begin{figure}[htp]
\centering
\includegraphics[scale=.5]{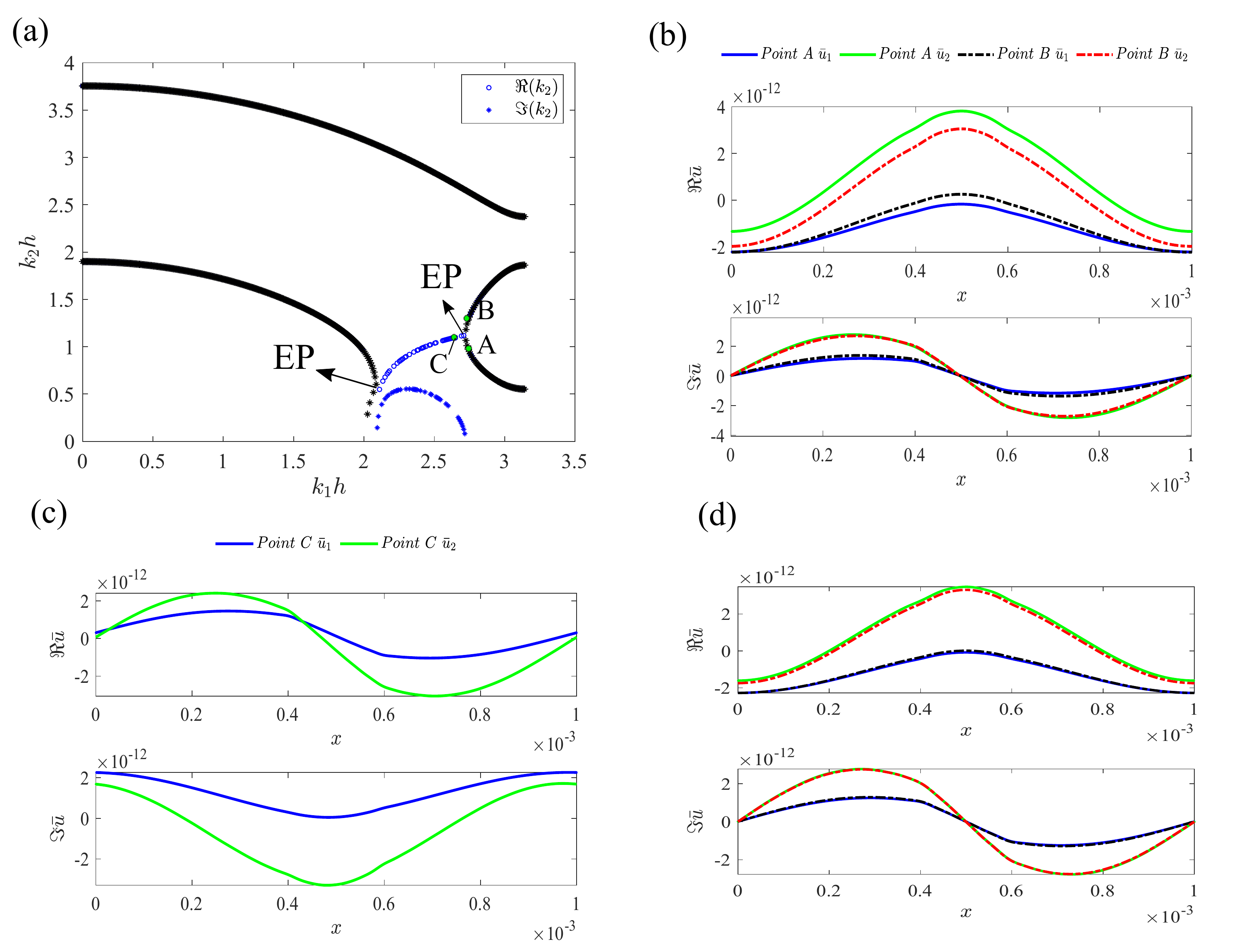}
\caption{Exceptional point at $f=1.80$ MHz. (a) The EP occurs at the intersection of a real branch and the branch with complex $k_2$ values (Blue curves). (b) The modeshapes corresponding to the points A and B on the right side of the EP. (c) The modeshapes corresponding to the points C on the complex $k_2$ branch (d) The modeshapes on the two points near the EP, which show the modeshape coalescence.}\label{fig-EPk2}
\end{figure}

Using the same layered composite as described earlier, and the $k_2(\omega,k_1)$ eigenvalue problem discussed in the previous section, the $k_1-k_2$ solutions calculated at $f=1.80$ MHz are shown in Fig. (\ref{fig-EPk2}a). In this figure, $k_1$ is taken to be real. The $k_2$ eigenvalues are of the following form: $a+ib,\ a-ib,\ -a+ib,\ -a-ib$. Only positive real branches and complex branches of the form $a+ib$ where $a,b>0$ are shown. There are two exceptional points in Fig. (\ref{fig-EPk2}a) which occur at the intersection of the complex $k_2$ branches (shown in blue color) and the real $k_2$ branches (shown in black). At these points, the imaginary parts of the complex branches go to zero and the real part equal the eigenvalues of the real branches. As these are the degenerecies of a non-Hermitian system, the corresponding eigenvectors usually coalesce. This is explicitly shown in Fig. (\ref{fig-EPk2}). The real and imaginary parts of the $\bar{u}_1$ and $\bar{u}_2$ modeshapes for different points close to the second EP at $k_1=2.71$ are plotted in Fig. (\ref{fig-EPk2}b-d). It can be seen in Fig. (\ref{fig-EPk2}d) that although the modeshapes are identical at the exceptional point, they diverge when nearby points are considered.  

\begin{figure}[htp]
\centering
\includegraphics[scale=.4]{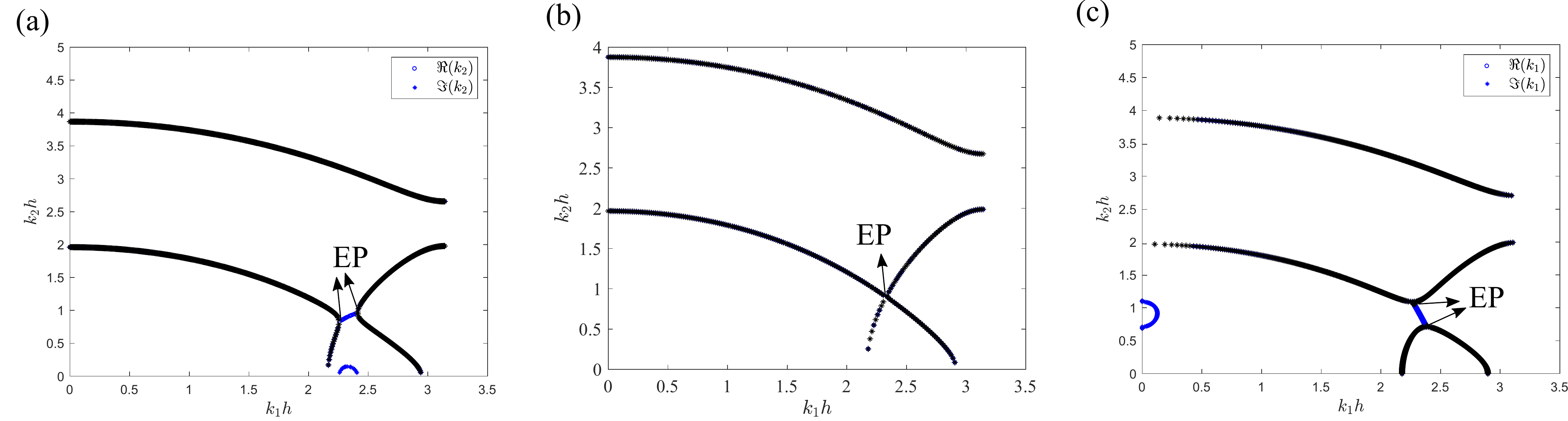}
\caption{Different exceptional points in the spectrum of the layered composite (a) EPs occuring at the intersections of a real branch and a complex $k_2$ branch at $f=1.858$ MHz. (b) EP occuring at the intersection of two real branches at $f=1.8621$ MHz. (c) EPs occuring at the intersection of a real branch and a complex $k_1$ branch at $f=1.864$ MHz }\label{realEP}
\end{figure}

The behavior of the exceptional points bears noting in this problem and is described in more detail in Fig. (\ref{realEP}). Fig. (\ref{realEP}a) shows a situation similar to Fig. (\ref{fig-EPk2}a) where two real branches are connected to each other by a pair of complex conjugate $k_2$ branches at a frequency of $1.858$ MHz. The intersections are exceptional points which are marked in the figure. As we increase the frequency, we note that the complex branches diminish and eventually disappear at around $1.862$ MHz. At this point, the two exceptional points coalesce in the real domain to give rise to a crossing of the propagating branches. The eigenvectors at this crossing point will still coalesce. Further increasing the frequency to $1.864$ MHz leads to a separation of the propagating modes but in the $k_2$ direction. Two exceptional points again emerge in this system but now they are at the intersections of the propagating modes with complex $k_1$ branches. This case is similar to what has been reported in \cite{Gal2019}.

\section{Solution of the scattering problem using Betti-Rayleigh reciprocity}

Now we consider the scattering problem at an interface between the laminate and homogeneous medium shown in Fig. (\ref{fschematic}). The incident, transmitted, and reflected fields are given in Eqs. (\ref{eIncident}) and (\ref{eScattered}). The goal is to calculate the scattering coefficients such that interface continuity and conservation of energy are satisfied. Ref. \cite{srivastava2017evanescent} has presented a solution to this problem by projecting the interface continuity conditions on to a suitable basis. We present an alternative solution strategy which depends upon the application of the Betti-Rayleigh reciprocity theorem \cite{Achenbach2004,achenbach2019} to the problem (see Appendix-B for more details on the formulation).

\subsection{Solving the Scattering Problem}\label{SEP}

\begin{figure}[htp]
\centering
\includegraphics[scale=.35]{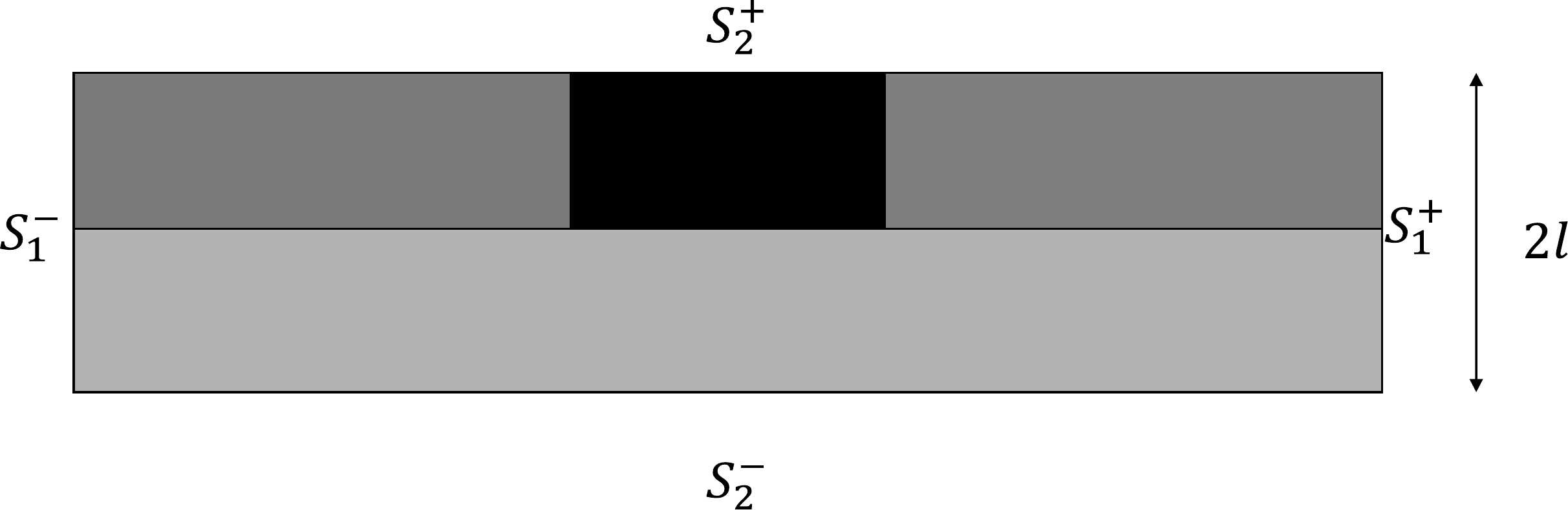}
\caption{Sub-domain considered for the application of Betti-Rayleigh reciprocity theorem.}\label{fdomain1}
\end{figure}

The incident, transmitted, and reflected fields are given in Eq. (\ref{eScattered}). For a longitudinal wave with wave number $k=k_l$ impinging the interface at an incident angle $\theta$ (with the normal), we have $\mathbf{d}=\pm[\sin(\theta),\cos(\theta)]^T$. Similarly, for transverse incident wave with wave number $k=k_t$, we have $\mathbf{d}=\pm[-\cos(\theta),\sin(\theta)]^T$. Lustig et. al. \cite{Gal2019} had to  characterized these reflected modes as longitudinal or transverse before using them in scattering calculations. An advantage of the eigenvalue problem of the form given in Eq. (\ref{mixed_form}) is that when it is used for a homogeneous medium, it automatically produces all the reflected modeshapes with correct polarization and directions of propagation. With this, the displacement ($u_1,u_2$) and traction continuity at the interface are given by:
\begin{eqnarray}
\label{bc1}
u_j^{I}+u_j^R=u_j^T\\
\label{bc2}
\sigma_{2j}^{I}+\sigma_{2j}^R=\sigma_{2j}^T
\end{eqnarray}
We seek to apply the reciprocity theorem on a finite sub-region of the infinite domain of Fig. (\ref{fschematic}). This finite sub-region, shown in Fig. (\ref{fdomain1}), is a rectangle of height $2l$ centered along $x_2=0$. The region encompasses one unit cell of the layered composite. For this configuration, the surface term in (\ref{eRec2}) is given by:
\begin{eqnarray}
\displaystyle \int_{\partial\Omega}(\cdot)dS=\int_{S_1^++S_1^-+S_2^++S_2^-}(\cdot)dS
\end{eqnarray}
Furthermore, we will consider the problem in the limit $l\rightarrow 0$ in which case the volume terms in Eq. (\ref{eRec2}) will go to zero. In this limit, $S_2^+$ corresponds to a surface just above the $x_2=0$ axis and $S_2^-$ corresponds to a surface just below it. These surfaces are represented by $0^+,0^-$ respectively. Reciprocity relation for the present problem is, therefore, given by:
\begin{eqnarray}\label{eRec3}
\nonumber
\displaystyle \int_{x_2=0^+}\left[t_{i}^Au_i^{B*}-t_{i}^Bu_i^{A*}\right]dS+\int_{x_2=0^-}\left[t_{i}^Au_i^{B*}-t_{i}^Bu_i^{A*}\right]dS=\\-\int_{S_1^-}\left[t_{i}^Au_i^{B*}-t_{i}^Bu_i^{A*}\right]dS-\int_{S_1^+}\left[t_{i}^Au_i^{B*}-t_{i}^Bu_i^{A*}\right]dS
\end{eqnarray}
We choose state $A$ to be the actual scattering problem under consideration with the reflected and transmitted fields given by Eqs. \ref{eIncident}-\ref{eScattered}. 
We consider virtual states $B$ from appropriate solutions of two problems: 
\begin{itemize}
    \item In-plane waves traveling in the homogeneous medium in Fig. (\ref{fschematic}) now extended infinitely. The waves are assumed to be harmonic at $\omega$ and with an $x_1$ wavenumber $k_1$ and $x_2$ wavenumber $\kappa_2$.
    \item In plane waves traveling in the layered medium in Fig. (\ref{fschematic}) now extended infinitely. The waves are assumed to be harmonic at $\omega$ and with an $x_1$ wavenumber $k_1$ and $x_2$ wavenumber $k_2$.
\end{itemize}
Each of these free virtual waves are feasible solutions in their respective infinite domain problems, and are easily determined from the eigenvalue formulation presented earlier. Since each of these virtual waves has an exponential term of the form $\exp(i(k_1x-\omega t))$ which they share with waves in state $A$, it can be shown that the only $x_1$ dependence in every integral in (\ref{eRec3}) comes from a unit cell periodic function. This further implies that the right hand side of Eq. (\ref{eRec3}) goes to zero for the set of chosen states B since $S_1^+$ and $S_1^-$ are one unit cell apart. For all such states, therefore, the reciprocity relation simplifies to:
\begin{eqnarray}\label{eRec4}
\displaystyle \int_{x_2=0^+}\left[t_{i}^Au_i^{B*}-t_{i}^Bu_i^{A*}\right]dS+\int_{x_2=0^-}\left[t_{i}^Au_i^{B*}-t_{i}^Bu_i^{A*}\right]dS=0
\end{eqnarray}

By considering all the $N_t+N_r$ virtual sates in the above equation and after some algebraic manipulations, the following linear system of equation is obtained (see Appendix-B for details):
\begin{equation}
\mathbf{PS}=\mathbf{I}
\end{equation}
where $\mathbf{P}$ is a square matrix of size $(N_t+N_r)\times(N_t+N_r)$ and $\mathbf{S}=[T^{(1)}\quad T^{(2)}\quad...\quad T^{(N_t)}\quad R^{(1)}\quad R^{(2)}\quad ...\quad R^{(N_r)}]^T$ is the scattering coefficients vector. The $\mathbf{P}$ matrix consists of inner products of the periodic components of displacement and stress modeshapes. These are automatically produced by the eigenvalue problem in Eq. (\ref{EStateSpace}). The total stress emerging from the solution of scattering problem is:
\begin{eqnarray}
\displaystyle \boldsymbol{\sigma}^T(x_1,x_2)=\sum_{i=1}^{N_t}T^{(i)}\bar{\boldsymbol{\sigma}}_T^{(i)}(x_1)\exp\left[i(k_1x_1+k_2^{(i)}x_2-\omega t)\right]\\
\boldsymbol{\sigma}^R(x_1,x_2)=\sum_{i=1}^{N_r}R^{(i)}\bar{\boldsymbol{\sigma}}_R^{(i)}(x_1)\exp\left[i(k_1x_1+\kappa_2^{(i)}x_2-\omega t)\right]
\end{eqnarray}

\subsection{Energy considerations and Biorthogonality}
As a check on the consistency of the calculations we also need to consider the balance of energy flow in the system. Consider a rectangular area of length $b$ and height $2l$ with its latter dimension bisected by $x_2=0$. In the absence of any dissipating mechanisms the total energy entering this rectangle should balance the energy leaving it. The time averaged real part of the Poynting vector gives the energy flux:
\begin{equation}
\boldsymbol{\mathcal{P}}=\pm\frac{1}{2}\Re{\left[\boldsymbol{\sigma}\cdot\dot{\mathbf{u}}^*\right]}
\end{equation}
(positive sign for reflected waves and negative sign for transmitted waves) which in the present case has two components $\mathcal{P}_1,\mathcal{P}_2$. The unit cell averages of these (denoted by the brackets $\langle \rangle$) give the time and unit cell averaged energy flux in a desired direction. The average energy entering this area due to the incident wave is $-0.5\Re{\left[i\omega(\sigma_{12}^Iu_1^{I*}+\sigma_{22}^Iu_2^{I*})\right]}$ where $\theta_i$ is the angle of incidence. There is a net loss of energy from this region due to the presence of transmitted and reflected waves. The energy contained in these waves is generated at the interface and then transported away and can be considered by using the field summation expressions introduced above. Ignoring the $\omega,k_1$ terms (they will cancel out), the Poynting vector in the transmitted and reflected regimes have the following form:
\begin{eqnarray}
\label{ePoynting}
\nonumber \displaystyle {\mathcal{P}}^T_2=\frac{1}{2}\Re{\left[-i\omega\hat{\mathbf{j}}\cdot\left(\sum_{i=1}^\infty T^{(i)}\bar{\boldsymbol{\sigma}}_T^{(i)}e^{ik_2^{(i)}l}\right)\cdot\left(\sum_{i=1}^\infty T^{(i)}\bar{\mathbf{u}}_T^{(i)}e^{ik_2^{(i)}l}\right)^* \right]}\\
\displaystyle {\mathcal{P}}^R_2=\frac{1}{2}\Re{\left[i\omega\hat{\mathbf{j}}\cdot\left(\sum_{i=1}^\infty R^{(i)}\bar{\boldsymbol{\sigma}}_R^{(i)}e^{-i\kappa_2^{(i)}l}\right)\cdot\left(\sum_{i=1}^\infty R^{(i)}\bar{\mathbf{u}}_R^{(i)}e^{-i\kappa_2^{(i)}l}\right)^* \right]}
\end{eqnarray}
The above terms have been evaluated at $x_2=l$ for the transmitted modes and at $x_2=-l$ for the reflected modes and they have an implicit dependence on $x_1$ since the modeshapes depend upon $x_1$. However, all $x_1$ dependent terms are, in fact, periodic with the unit cell. Therefore, taking their unit cell averages removes the $x_1$ dependence. We denote the unit cell averages with $\langle \boldsymbol{\mathcal{P}}^T\rangle,\langle\boldsymbol{\mathcal{P}}^R\rangle$. The energy leaving the region of interest due to the transmitted waves is now given by $b\langle{\mathcal{P}}^T\rangle_2$ and due to the reflected waves is given by $b\langle{\mathcal{P}}^R\rangle_2$. Energy balance equation thus becomes:
\begin{eqnarray}
\label{energyidentity}
\displaystyle E=\frac{2(\langle{\mathcal{P}}^T\rangle_2+\langle{\mathcal{P}}^R\rangle_2)}{\Re{\left[-i\omega(\sigma_{12}^Iu_1^{I*}+\sigma_{22}^Iu_2^{I*})\right]}}=1
\end{eqnarray}
Furthermore, we can calculate the Poynting vector for each mode:
\begin{eqnarray}
\displaystyle \mathcal{P}^{(i)}=\pm\frac{1}{2}\mathcal{R}\left[\bar{\boldsymbol{\sigma}}^{(i)}\cdot \dot{\bar{\mathbf{u}}}^{(i)*}\right]
\end{eqnarray}
which then gives us the refraction angle of each mode:
\begin{eqnarray}
\displaystyle \theta^{(i)}=\arctan(\frac{\langle\mathcal{P}^{(i)}_1\rangle}{\langle\mathcal{P}^{(i)}_2\rangle})
\end{eqnarray}
One curious aspect of the energy flux conservation above is that it needs to be evaluated at some $x_2$ value. This is due to the fact that in the unit cell averages of Eq. (\ref{ePoynting}), the double summation cannot be reduced to a single summation through some orthogonality condition. The exponential terms, therefore, cannot cancel out and there remains a dependence on $x_2$. Strictly speaking, therefore, the energy balance appears not to be independent of $x_2$. This means that potentially the Poynting vector terms calculated at some other $x_2$ values in Eq. (\ref{ePoynting}) may lead to a violation of Eq. (\ref{energyidentity}) -- we want to emphasize that we do not see this violation in numerical investigations, however, the point still remains that energy flux identity written above has an apparent $x_2$ dependence. To rectify this situation, we investigate an alternative formulation for the energy metric by considering biorthogonality inherent in the problem. From \cite{mokhtari2019properties}, the biorthogonality relationship (after appropriate normalization) is:
\begin{eqnarray}
\displaystyle \langle B\phi^{(p)}_r,\phi^{(q)}_l\rangle=\delta_{pq}
\end{eqnarray}
where $\phi^{(p)}_r,\phi^{(q)}_l$ are the right and left eigenvectors corresponding to the $p^\mathrm{th},q^\mathrm{th}$ eigenvalues respectively. For the present problem, the above equation in the transmitted regime, as an example, is \cite{mokhtari2019properties}:
\begin{eqnarray}
\displaystyle \langle n_j\bar{\sigma}^{(p)}_{ij(r)},\bar{u}^{(q)}_{i(l)} \rangle-\langle n_j\bar{u}^{(p)}_{i(r)},\bar{\sigma}^{(q)}_{ij(l)} \rangle=\delta_{pq}
\end{eqnarray}
The above equation is general but in the present case $\mathbf{n}$ would be the unit vector in the $x_2$ direction ($\mathbf{n}=\hat{\mathbf{j}}$). We have omitted the subscript $T$ to make the notation easier. The unit cell averaged energy flux in the direction of $\mathbf{n}$ may be derived from the Poynting vector as:
\begin{eqnarray}
\displaystyle E_n=-\frac{1}{2}\langle\Re{\left[n_j\sigma_{ij}\dot{u}^*_i\right]}\rangle=\frac{\omega}{2}\langle\Re{\left[in_j\sigma_{ij}{u}^*_i\right]}\rangle=-\frac{\omega}{2}\Im\langle{\left[n_j\sigma_{ij}{u}^*_i\right]}\rangle=\frac{i\omega}{4}\left[\langle{n_j\sigma_{ij},{u}_i}\rangle-\langle{n_j{u}_i,\sigma_{ij}}\rangle\right]
\end{eqnarray}
Now the main idea is to expand stress and displacement fields in terms of the left and right eigenvectors. On the transmitted side, we can write:
\begin{eqnarray}
\displaystyle \nonumber \mathbf{u}=\sum_{i=1}^{\infty}T_r^{(i)}\bar{\mathbf{u}}^{(i)}_re^{ik_2^{(i)}l}=\sum_{i=1}^{\infty}T_l^{(i)}\bar{\mathbf{u}}^{(i)}_le^{ik_2^{(i)}l}\\
\displaystyle \boldsymbol{\sigma}=\sum_{i=1}^{\infty}T_r^{(i)}\bar{\boldsymbol{\sigma}}^{(i)}_re^{ik_2^{(i)}l}=\sum_{i=1}^{\infty}T_l^{(i)}\bar{\boldsymbol{\sigma}}^{(i)}_le^{ik_2^{(i)}l}
\end{eqnarray}
where $\bar{\mathbf{u}}_r,\bar{\mathbf{u}}_l,\bar{\boldsymbol{\sigma}}_r,\bar{\boldsymbol{\sigma}}_l$ are extracted from the right and left eigenvectors respectively and $T_r,T_l$ are the corresponding scattering coefficients. Similar expressions hold on the reflection side as well. Since $\mathbf{n}=\hat{\mathbf{j}}$, the 2-component of energy flux on the transmitted side is:
\begin{eqnarray}
\displaystyle E_2^T=\frac{i\omega}{4}\left[\langle\sum_{p}T^{(p)}_rn_j\bar{\sigma}^{(p)}_{ij,r}e^{ik_2^{(p)}l},\sum_{q}T^{(q)}_l\bar{u}^{(q)}_{i,l}e^{ik_2^{(q)}l}\rangle-\langle\sum_{p}T^{(p)}_rn_j\bar{u}^{(p)}_{i,r}e^{ik_2^{(p)}l},\sum_{q}T^{(q)}_l\bar{\sigma}^{(q)}_{ij,l}e^{ik_2^{(q)}l}\rangle\right]
\end{eqnarray}
An arbitrary term in the above double summation is:
\begin{eqnarray}
\displaystyle \frac{i\omega T^{(p)}_rT^{(q*)}_l}{4}\left[\langle n_j\bar{\sigma}^{(p)}_{ij,r},\bar{u}^{(q)}_{i,l}\rangle-\langle n_j\bar{u}^{(p)}_{i,r},\bar{\sigma}^{(q)}_{ij,l}\rangle\right]e^{i(k_2^{(p)}-k_2^{(q)})l}=-\frac{i\omega T^{(p)}_rT^{(q*)}_l}{4}e^{i(k_2^{(p)}-k_2^{(q)})l}\delta_{pq}
\end{eqnarray}
Therefore, the 2-component of energy flux on the transmitted side is:
\begin{eqnarray}
\displaystyle E_2^T=\frac{i\omega}{4}\sum_{p}T^{(p)}_rT^{(p*)}_l
\end{eqnarray}
Similarly, the 2-component of energy flux on the reflected side is:
\begin{eqnarray}
\displaystyle E_2^R=\frac{i\omega}{4}\sum_{p}R^{(p)}_rR^{(p*)}_l
\end{eqnarray}
showing that both energy fluxes are, in fact, independent of $x_2$.

\section{Numerical examples}
\begin{figure}[htp]
\centering
\includegraphics[scale=.55]{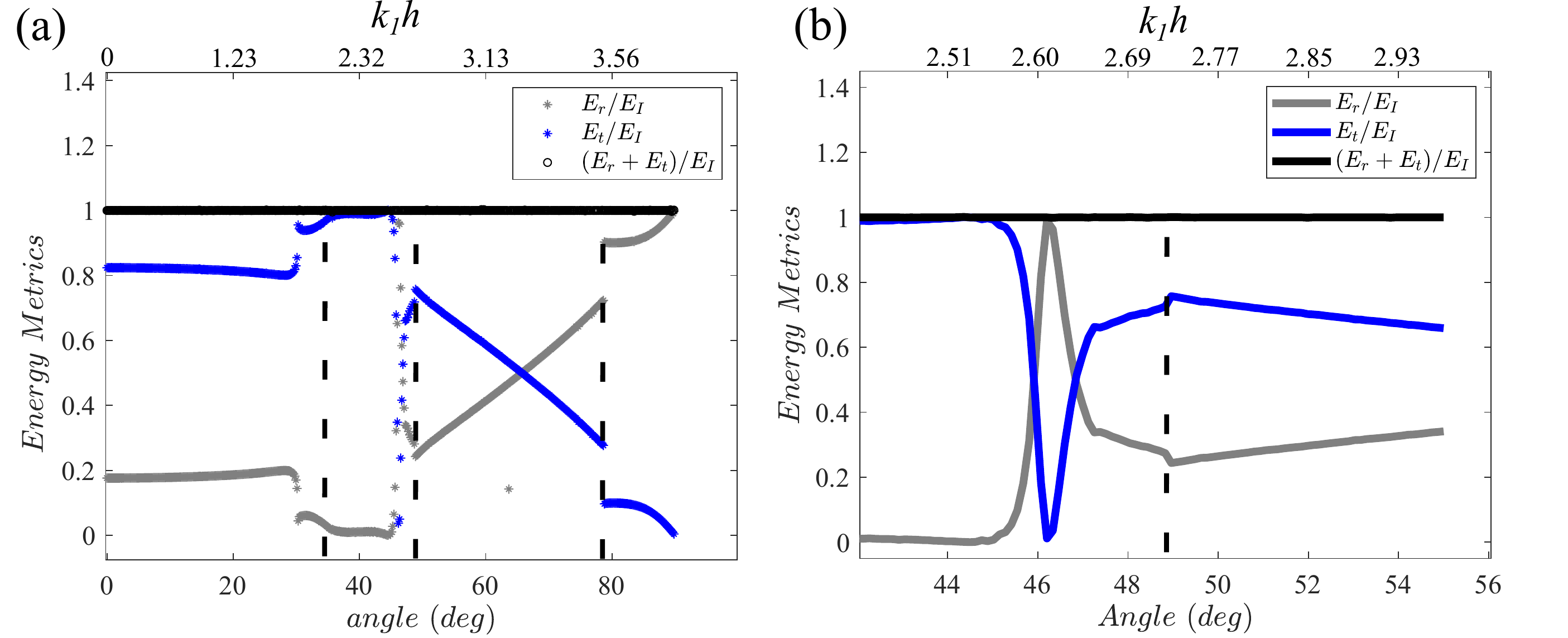}
\caption{(a) Energy balance for the scattering problem at $f=1.80$ MHz. (b) Zoomed region for angles $42^{\circ}-56^{\circ}$}\label{energy180}. The locations of EPs are indicated with dashed lines.
\end{figure}
In this section we solve some scattering examples based on the previous sections. Consider an interface between a 2-phase layered composite and a homogeneous medium as shown in Fig. (\ref{fschematic}). The material properties of the homogeneous medium are assumed to be: $\lambda_h=51.1$ GPa, $\mu_h=26.3$ GPa, $\rho_h=2700$ kg/m$^3$ and for the layered composite: $\lambda_1=121.1$ GPa, $\mu_1=80.8$ GPa, $\rho_1=7800$ kg/m$^3$, $h_1=0.8$ mm and $\lambda_2=51.1$ GPa, $\mu_2=26.3$ GPa, $\rho_2=2700$ kg/m$^3$, and $h_2=0.2$ mm. A shear in-plane wave in the homogeneous material is incident at the interface with an incidence angle $\theta$ (measured from the normal to the interface). The $k_1=\omega/\sqrt{\mu_h/\rho_h}\sin\theta$ component of the wavevector is known and is conserved across the interface due to Snell's law. Knowing the frequency and $k_1$, we find all the $k_2$ components of the wavevector and their corresponding displacement and stress modeshapes by solving the $k_2(\omega, k_1)$ eigenvalue problem characterized by Eq. (\ref{Eq:beta3problem}). The $k_2$ values in the homogeneous medium are also calculated from the linear non self-adjoin form or the eigenvalue problem. As mentioned earlier, not all $k_2$ values are admissible for inclusion in the summation terms. Admissible $k_2$ values are chosen with the requirement that either their flux in the $x_2$ direction is 0 or going away from the interface. Knowing all the admissible $k_2$ values and the corresponding displacement and stress modeshapes, the Betti-Rayleigh reciprocity theorem is applied to find the scattering coefficients $T^{(i)}$ and $R^{(i)}$ in Eq. (\ref{eScattered}). The displacement and stress field can be completely determined as they are expanded according to Eq. (\ref{eScattered})
once the scattering coefficients are known. To ensure the correctness of the solution, the energy flux consistency in the $x_2$ direction is checked by applying Eq. (\ref{energyidentity}). 
\begin{figure}[htp]
\centering
\includegraphics[scale=.55]{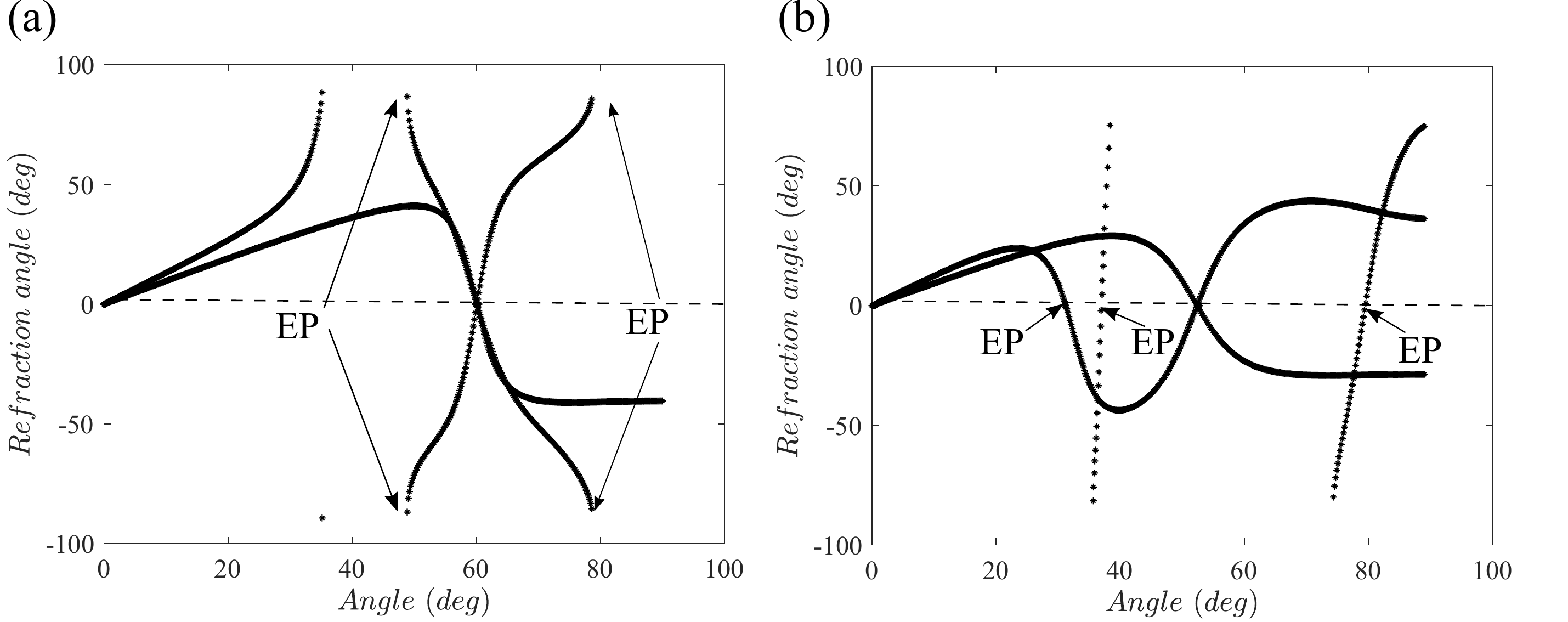}
\caption{Refraction angles of the propagating transmitted waves at (a) $f=1.80$ MHz and (b) $f=1.97$} MHz. The $x$ axis represents the angle of incidence of the incident wave.\label{refraction}
\end{figure}

The first example is for a shear in-plane wave at frequency $f=1.80$ MHz for which the dispersion plots have been discussed in Fig. (\ref{fig-EPk2}). The energy metric plots are shown in Fig. (\ref{energy180}a). The blue curve represents total normalized transmitted energy flux in the $x_2$ direction and the grey curve represents total normalized reflected energy flux. The black curve is the sum of the two. Since the black curve equals 1 for all angles of incidences, the energy flux is satisfied almost exactly in the scattering calculations giving a high degree of confidence in the results. In the incidence angle region of consideration, there are three exceptional points. These are at $\theta\approx 35.4^{\circ}, 48.6^{\circ}$, and $78.4^{\circ}$ with $\theta\approx78.4^{\circ}$ being in the second Brillouin zone and they are marked in Fig. (\ref{energy180}a). As these EPs are associated with either the emergence or annihilation of branches as one changes the incidence angle in their vicinity (Fig. (\ref{fig-EPk2})), the scattered energy in the vicinity of these EPs exhibits fast variations. Another interesting phenonmenon in Fig. (\ref{energy180}a) is the sudden drop in transmitted energy just above 46 degrees (shown in more detail in Fig. (\ref{energy180}b)). This region corresponds to $k_1h\approx 2.5$ and referring to Fig. (\ref{fig-EPk2}a), it is clear that there is only one propagating mode in this region and there are no EPs in the immediate vicinity. This sudden drop of energy is, therefore, unexplained merely from arguments of EPs or emerging/disappearing branches and may be related to exotic wave effects like resonance trapping\cite{Mller2009,Persson2000}. It is clear that this sudden drop of energy is not obvious from the bandstructure and would be completely missed without conducting scattering calculations. A further interesting aspect of this problem is that wave directions seem to flip signs abruptly across an EP. For $f=1.80$ MHz, Fig. (\ref{refraction}a) shows refraction angle calculations for various propagating modes as a function of the incidence angle. These calculations are based upon Poynting vector calculations and are representative of group velocity directions. Referring back to Fig. (\ref{fig-EPk2}a), we note that there are two propagating branches to the left of the first EP which disappear to the right, and two propagating branches to the right of the second EP which disappear to the left. Fig. (\ref{refraction}a) clearly shows that the two propagating branches, in both cases, refract in opposite directions. 

\begin{figure}[htp]
\centering
\includegraphics[scale=.25]{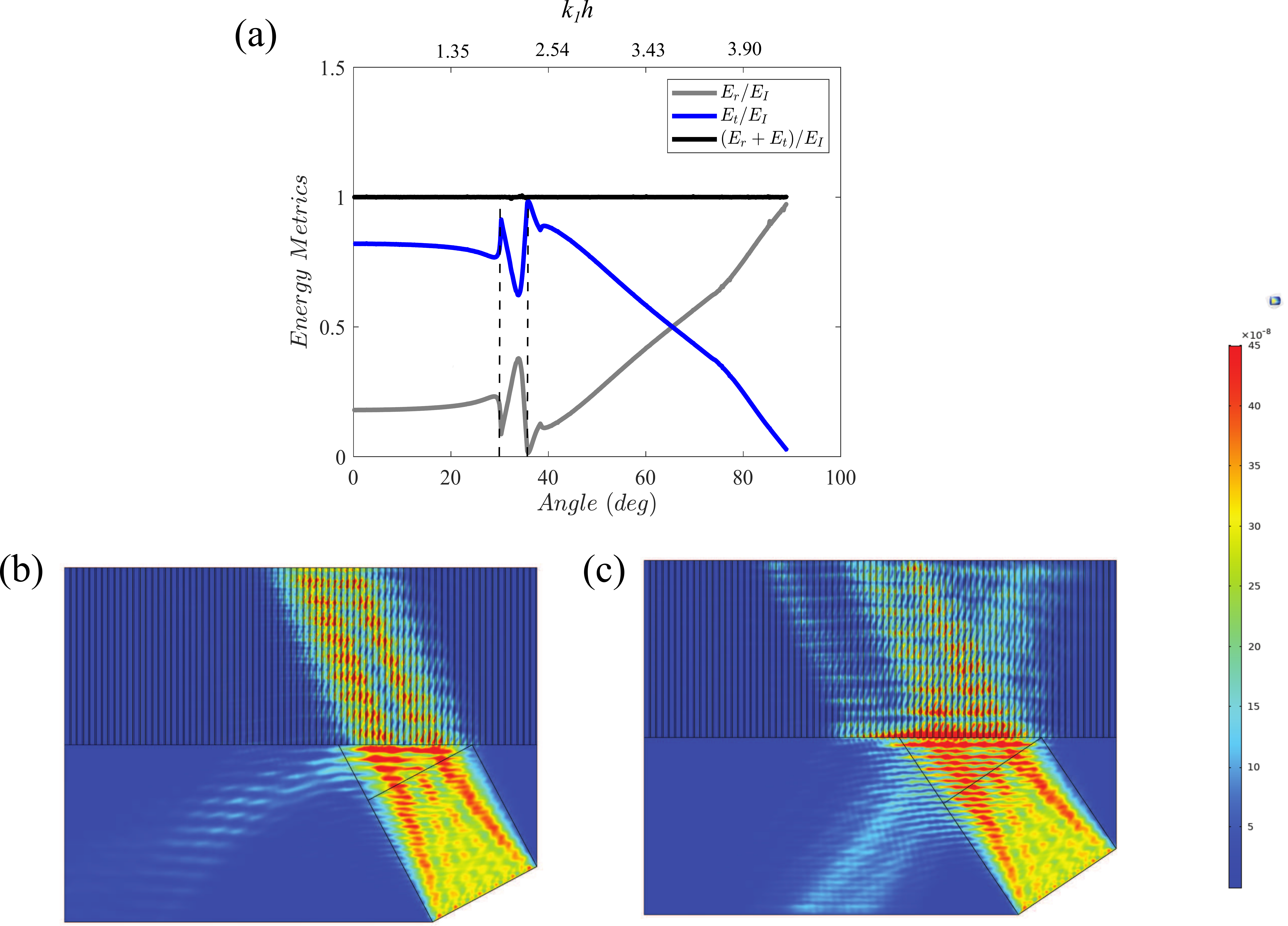}
\caption{ Energy metrics for $f=1.97$ MHz are shown in (a). Dashed lines are at the locations of the  EPs in the first Brillouin zone. (b) Elastic strain and kinetic energy density at $f=1.97$ MHz and $\theta=28^{\circ}$ (before the EP) where there is mode with positive refraction and (c)  $\theta=34^{\circ}$ (after the EP) where there is negative refraction. The first EP can be reached at $\theta=31.5^{\circ}$. }\label{FEMenergy197}
\end{figure}

This effect is also present at other EPs. We give another set of examples in Fig. (\ref{refraction}b) where the refraction angles are plotted for the propagating modes at 1.97 MHz. Referring to Fig. (\ref{k1EP}a), we note that at this frequency there are two EPs present in the first Brillouin zone and that these EPs are in the $k_1$ domain as opposed to the EPs at 1.80 MHz which are in the $k_2$ domain. Fig. (\ref{refraction}b) shows refraction angle sign flips at three EPs. The first two EPs are in the first Brillouin zone whereas the third EP is in the second Brillouin zone. Regardless, there exist sign flips for all three EPs. Fig. (\ref{FEMenergy197}a) shows the energy metrics for this case showing the locations of the EPs and the corresponding changes in the scattered energy in the vicinity of these EPs. Figs. (\ref{FEMenergy197}b, c) show FEM calculations exploring the scattered field at 1.97 MHz at two different angles of incidence. The FEM model contains 75 repeated unit cells with material properties and dimensions described in the beginning of this section. The incident wave is excited as a line source on a boundary located on the homogeneous domain. Absorbing boundary conditions are applied at all other external boundaries of the simulation domain to eliminate effects of reflections that would otherwise be present at the boundaries. Fig. (\ref{FEMenergy197}b) shows FEM simulation at an incidence angle of 28 degrees and Fig. (\ref{FEMenergy197}c) shows the simulation at 34 degrees. These two angles lie on two sides of the left EP (which lies at 31.5 degree incidence) in Fig. (\ref{k1EP}a) and both angles lie to the left of the lowest branch. We note that while there is no negative refraction at 28 degrees, there is a negatively refracted mode at 34 degrees. We also note that there is no new branch being introduced at 34 degrees since the third branch begins at a higher angle of incidence. Therefore, the negatively refracted signal must be coming from the lower of the two propagating branches which has transitioned through the first and, thus, has transitioned from a positively refracting to a negatively refracting branch (see Fig. \ref{refraction}b).

\section{Conclusions}

We considered the problem of the scattering of in-plane waves at an interface between a homogeneous medium and a layered composite. The relevant eigenmodes in the two regions come from non self-adjoint eigenvalue operators and are calculated by solving a recently described non self-adjoint eigenvalue problem particularly suited to scattering studies \cite{mokhtari2019properties}. In a model composite, we elucidate the emergence of a rich spectrum of eigenvalue degeneracies. These degeneracies appear in both the complex and real domains of the wave-vector. However, since this problem is non self-adjoint, these degeneracies generally represent a coalescing of both the eigenvalues and eigenvectors (exceptional points). The presence of these degeneracies, in some cases, hints at fast changes in the scattered field as the incident angle is changed by small amounts. In all the cases considered in this paper, these EPs also represent points in the spectrum where a wave flips the direction of its refraction from positive to negative or vice-versa as the EPs are traversed. We calculate the scattered fields through a novel application of the Betti-Rayleigh reciprocity theorem. We also show that there exists a deep between energy flux conservation and the biorthogonality relationship of the non self-adjoint problem. This allows us to diagonalize the flux conservation relationship, elucidating the interplay between the scattering coefficients of the left and right eigenvectors. We note the important result that such diagonalization is impossible without the use of biorthogonality. We have presented several numerical examples showing a rich scattering spectrum. This includes demonstrating fast variations in the scattered energy in the vicinity of EPs and refraction angle sign flip across the EPs. In one particularly intriguing example, we point out wave behavior which may be related to the phenomenon of resonance trapping - hitherto unknown in the present class of problems. This phenomenon along with the refraction angle sign flips are not obvious merely from the bandstructure of the problem and require the calculation of Poynting vectors and scattered fields.

\acknowledgments     
 
A.S. acknowledges support from the NSF CAREER grant \#1554033 to the Illinois Institute of Technology and NSF grant \#1825354 to the Illinois Institute of Technology. A.V.A. acknowledges NSF grant \#1825969 to the University of Massachusetts, Lowell.


\begin{thebibliography}{66}%
\makeatletter
\providecommand \@ifxundefined [1]{%
 \@ifx{#1\undefined}
}%
\providecommand \@ifnum [1]{%
 \ifnum #1\expandafter \@firstoftwo
 \else \expandafter \@secondoftwo
 \fi
}%
\providecommand \@ifx [1]{%
 \ifx #1\expandafter \@firstoftwo
 \else \expandafter \@secondoftwo
 \fi
}%
\providecommand \natexlab [1]{#1}%
\providecommand \enquote  [1]{``#1''}%
\providecommand \bibnamefont  [1]{#1}%
\providecommand \bibfnamefont [1]{#1}%
\providecommand \citenamefont [1]{#1}%
\providecommand \href@noop [0]{\@secondoftwo}%
\providecommand \href [0]{\begingroup \@sanitize@url \@href}%
\providecommand \@href[1]{\@@startlink{#1}\@@href}%
\providecommand \@@href[1]{\endgroup#1\@@endlink}%
\providecommand \@sanitize@url [0]{\catcode `\\12\catcode `\$12\catcode
  `\&12\catcode `\#12\catcode `\^12\catcode `\_12\catcode `\%12\relax}%
\providecommand \@@startlink[1]{}%
\providecommand \@@endlink[0]{}%
\providecommand \url  [0]{\begingroup\@sanitize@url \@url }%
\providecommand \@url [1]{\endgroup\@href {#1}{\urlprefix }}%
\providecommand \urlprefix  [0]{URL }%
\providecommand \Eprint [0]{\href }%
\providecommand \doibase [0]{http://dx.doi.org/}%
\providecommand \selectlanguage [0]{\@gobble}%
\providecommand \bibinfo  [0]{\@secondoftwo}%
\providecommand \bibfield  [0]{\@secondoftwo}%
\providecommand \translation [1]{[#1]}%
\providecommand \BibitemOpen [0]{}%
\providecommand \bibitemStop [0]{}%
\providecommand \bibitemNoStop [0]{.\EOS\space}%
\providecommand \EOS [0]{\spacefactor3000\relax}%
\providecommand \BibitemShut  [1]{\csname bibitem#1\endcsname}%
\let\auto@bib@innerbib\@empty
\bibitem [{\citenamefont {Nemat-Nasser}(2015{\natexlab{a}})}]{nemat2015anti}%
  \BibitemOpen
  \bibfield  {author} {\bibinfo {author} {\bibfnamefont {Sia}\ \bibnamefont
  {Nemat-Nasser}},\ }\bibfield  {title} {\enquote {\bibinfo {title} {Anti-plane
  shear waves in periodic elastic composites: band structure and anomalous wave
  refraction},}\ }in\ \href {\doibase 10.1098/rspa.2015.0152} {\emph {\bibinfo
  {booktitle} {Proc. R. Soc. A}}},\ Vol.\ \bibinfo {volume} {471}\ (\bibinfo
  {organization} {The Royal Society},\ \bibinfo {year} {2015})\BibitemShut
  {NoStop}%
\bibitem [{\citenamefont {Shmuel}\ and\ \citenamefont
  {Band}(2016)}]{shmuel2016universality}%
  \BibitemOpen
  \bibfield  {author} {\bibinfo {author} {\bibfnamefont {Gal}\ \bibnamefont
  {Shmuel}}\ and\ \bibinfo {author} {\bibfnamefont {Ram}\ \bibnamefont
  {Band}},\ }\bibfield  {title} {\enquote {\bibinfo {title} {Universality of
  the frequency spectrum of laminates},}\ }\href@noop {} {\bibfield  {journal}
  {\bibinfo  {journal} {Journal of the Mechanics and Physics of Solids}\
  }\textbf {\bibinfo {volume} {92}},\ \bibinfo {pages} {127--136} (\bibinfo
  {year} {2016})}\BibitemShut {NoStop}%
\bibitem [{\citenamefont {Chen}\ and\ \citenamefont
  {Elbanna}(2016)}]{chen2016modulating}%
  \BibitemOpen
  \bibfield  {author} {\bibinfo {author} {\bibfnamefont {Qianli}\ \bibnamefont
  {Chen}}\ and\ \bibinfo {author} {\bibfnamefont {Ahmed}\ \bibnamefont
  {Elbanna}},\ }\bibfield  {title} {\enquote {\bibinfo {title} {Modulating
  elastic band gap structure in layered soft composites using sacrificial
  interfaces},}\ }\href@noop {} {\bibfield  {journal} {\bibinfo  {journal}
  {Journal of Applied Mechanics}\ }\textbf {\bibinfo {volume} {83}},\ \bibinfo
  {pages} {111009} (\bibinfo {year} {2016})}\BibitemShut {NoStop}%
\bibitem [{\citenamefont {Mokhtari}\ \emph
  {et~al.}(2019{\natexlab{a}})\citenamefont {Mokhtari}, \citenamefont {Lu},\
  and\ \citenamefont {Srivastava}}]{MOKHTARI2019256}%
  \BibitemOpen
  \bibfield  {author} {\bibinfo {author} {\bibfnamefont {Amir~Ashkan}\
  \bibnamefont {Mokhtari}}, \bibinfo {author} {\bibfnamefont {Yan}\
  \bibnamefont {Lu}}, \ and\ \bibinfo {author} {\bibfnamefont {Ankit}\
  \bibnamefont {Srivastava}},\ }\bibfield  {title} {\enquote {\bibinfo {title}
  {On the emergence of negative effective density and modulus in 2-phase
  phononic crystals},}\ }\href {\doibase
  https://doi.org/10.1016/j.jmps.2019.02.016} {\bibfield  {journal} {\bibinfo
  {journal} {Journal of the Mechanics and Physics of Solids}\ }\textbf
  {\bibinfo {volume} {126}},\ \bibinfo {pages} {256 -- 271} (\bibinfo {year}
  {2019}{\natexlab{a}})}\BibitemShut {NoStop}%
\bibitem [{\citenamefont {Amirkhizi}\ and\ \citenamefont
  {Alizadeh}(2018)}]{Amirkhizi2018}%
  \BibitemOpen
  \bibfield  {author} {\bibinfo {author} {\bibfnamefont {Alireza~V.}\
  \bibnamefont {Amirkhizi}}\ and\ \bibinfo {author} {\bibfnamefont {Vahidreza}\
  \bibnamefont {Alizadeh}},\ }\bibfield  {title} {\enquote {\bibinfo {title}
  {Overall constitutive description of symmetric layered media based on
  scattering of oblique {SH} waves},}\ }\href {\doibase
  10.1016/j.wavemoti.2018.10.001} {\bibfield  {journal} {\bibinfo  {journal}
  {Wave Motion}\ }\textbf {\bibinfo {volume} {83}},\ \bibinfo {pages}
  {214--226} (\bibinfo {year} {2018})}\BibitemShut {NoStop}%
\bibitem [{\citenamefont {Hajarolasvadi}\ and\ \citenamefont
  {Elbanna}(2019)}]{Hajarolasvadi2019}%
  \BibitemOpen
  \bibfield  {author} {\bibinfo {author} {\bibfnamefont {Setare}\ \bibnamefont
  {Hajarolasvadi}}\ and\ \bibinfo {author} {\bibfnamefont {Ahmed~E}\
  \bibnamefont {Elbanna}},\ }\bibfield  {title} {\enquote {\bibinfo {title}
  {Dynamics of metamaterial beams consisting of periodically-coupled parallel
  flexural elements: a theoretical study},}\ }\href {\doibase
  10.1088/1361-6463/ab1f9e} {\bibfield  {journal} {\bibinfo  {journal} {Journal
  of Physics D: Applied Physics}\ }\textbf {\bibinfo {volume} {52}},\ \bibinfo
  {pages} {315101} (\bibinfo {year} {2019})}\BibitemShut {NoStop}%
\bibitem [{\citenamefont {Nejadsadeghi}\ \emph {et~al.}(2019)\citenamefont
  {Nejadsadeghi}, \citenamefont {Placidi}, \citenamefont {Romeo},\ and\
  \citenamefont {Misra}}]{Nejadsadeghi2019}%
  \BibitemOpen
  \bibfield  {author} {\bibinfo {author} {\bibfnamefont {Nima}\ \bibnamefont
  {Nejadsadeghi}}, \bibinfo {author} {\bibfnamefont {Luca}\ \bibnamefont
  {Placidi}}, \bibinfo {author} {\bibfnamefont {Maurizio}\ \bibnamefont
  {Romeo}}, \ and\ \bibinfo {author} {\bibfnamefont {Anil}\ \bibnamefont
  {Misra}},\ }\bibfield  {title} {\enquote {\bibinfo {title} {Frequency band
  gaps in dielectric granular metamaterials modulated by electric field},}\
  }\href {\doibase 10.1016/j.mechrescom.2019.01.006} {\bibfield  {journal}
  {\bibinfo  {journal} {Mechanics Research Communications}\ }\textbf {\bibinfo
  {volume} {95}},\ \bibinfo {pages} {96--103} (\bibinfo {year}
  {2019})}\BibitemShut {NoStop}%
\bibitem [{\citenamefont {Misra}\ and\ \citenamefont
  {Nejadsadeghi}(2019)}]{Misra2019}%
  \BibitemOpen
  \bibfield  {author} {\bibinfo {author} {\bibfnamefont {Anil}\ \bibnamefont
  {Misra}}\ and\ \bibinfo {author} {\bibfnamefont {Nima}\ \bibnamefont
  {Nejadsadeghi}},\ }\bibfield  {title} {\enquote {\bibinfo {title}
  {Longitudinal and transverse elastic waves in 1d granular materials modeled
  as micromorphic continua},}\ }\href {\doibase 10.1016/j.wavemoti.2019.05.005}
  {\bibfield  {journal} {\bibinfo  {journal} {Wave Motion}\ }\textbf {\bibinfo
  {volume} {90}},\ \bibinfo {pages} {175--195} (\bibinfo {year}
  {2019})}\BibitemShut {NoStop}%
\bibitem [{\citenamefont {Huang}\ \emph {et~al.}(2009)\citenamefont {Huang},
  \citenamefont {Sun},\ and\ \citenamefont {Huang}}]{huang2009negative}%
  \BibitemOpen
  \bibfield  {author} {\bibinfo {author} {\bibfnamefont {HH}~\bibnamefont
  {Huang}}, \bibinfo {author} {\bibfnamefont {CT}~\bibnamefont {Sun}}, \ and\
  \bibinfo {author} {\bibfnamefont {GL}~\bibnamefont {Huang}},\ }\bibfield
  {title} {\enquote {\bibinfo {title} {On the negative effective mass density
  in acoustic metamaterials},}\ }\href@noop {} {\bibfield  {journal} {\bibinfo
  {journal} {International Journal of Engineering Science}\ }\textbf {\bibinfo
  {volume} {47}},\ \bibinfo {pages} {610--617} (\bibinfo {year}
  {2009})}\BibitemShut {NoStop}%
\bibitem [{\citenamefont {Zhu}\ \emph {et~al.}(2014{\natexlab{a}})\citenamefont
  {Zhu}, \citenamefont {Liu}, \citenamefont {Hu}, \citenamefont {Sun},\ and\
  \citenamefont {Huang}}]{Huang2014}%
  \BibitemOpen
  \bibfield  {author} {\bibinfo {author} {\bibfnamefont {R.}~\bibnamefont
  {Zhu}}, \bibinfo {author} {\bibfnamefont {X.~N.}\ \bibnamefont {Liu}},
  \bibinfo {author} {\bibfnamefont {G.~K.}\ \bibnamefont {Hu}}, \bibinfo
  {author} {\bibfnamefont {C.~T.}\ \bibnamefont {Sun}}, \ and\ \bibinfo
  {author} {\bibfnamefont {G.~L.}\ \bibnamefont {Huang}},\ }\bibfield  {title}
  {\enquote {\bibinfo {title} {Negative refraction of elastic waves at the
  deep-subwavelength scale in a single-phase metamaterial},}\ }\href
  {https://doi.org/10.1038/ncomms6510} {\bibfield  {journal} {\bibinfo
  {journal} {Nature Communications}\ }\textbf {\bibinfo {volume} {5}},\
  \bibinfo {pages} {5510 EP --} (\bibinfo {year} {2014}{\natexlab{a}})},\
  \bibinfo {note} {article}\BibitemShut {NoStop}%
\bibitem [{\citenamefont {Liu}\ \emph {et~al.}(2011)\citenamefont {Liu},
  \citenamefont {Hu}, \citenamefont {Huang},\ and\ \citenamefont
  {Sun}}]{Liu2011}%
  \BibitemOpen
  \bibfield  {author} {\bibinfo {author} {\bibfnamefont {X.~N.}\ \bibnamefont
  {Liu}}, \bibinfo {author} {\bibfnamefont {G.~K.}\ \bibnamefont {Hu}},
  \bibinfo {author} {\bibfnamefont {G.~L.}\ \bibnamefont {Huang}}, \ and\
  \bibinfo {author} {\bibfnamefont {C.~T.}\ \bibnamefont {Sun}},\ }\bibfield
  {title} {\enquote {\bibinfo {title} {An elastic metamaterial with
  simultaneously negative mass density and bulk modulus},}\ }\href {\doibase
  10.1063/1.3597651} {\bibfield  {journal} {\bibinfo  {journal} {Applied
  Physics Letters}\ }\textbf {\bibinfo {volume} {98}},\ \bibinfo {pages}
  {251907} (\bibinfo {year} {2011})}\BibitemShut {NoStop}%
\bibitem [{\citenamefont {Hussein}\ \emph {et~al.}(2014)\citenamefont
  {Hussein}, \citenamefont {Leamy},\ and\ \citenamefont
  {Ruzzene}}]{hussein2014dynamics}%
  \BibitemOpen
  \bibfield  {author} {\bibinfo {author} {\bibfnamefont {Mahmoud~I}\
  \bibnamefont {Hussein}}, \bibinfo {author} {\bibfnamefont {Michael~J}\
  \bibnamefont {Leamy}}, \ and\ \bibinfo {author} {\bibfnamefont {Massimo}\
  \bibnamefont {Ruzzene}},\ }\bibfield  {title} {\enquote {\bibinfo {title}
  {Dynamics of phononic materials and structures: Historical origins, recent
  progress, and future outlook},}\ }\href@noop {} {\bibfield  {journal}
  {\bibinfo  {journal} {Applied Mechanics Reviews}\ }\textbf {\bibinfo {volume}
  {66}},\ \bibinfo {pages} {040802} (\bibinfo {year} {2014})}\BibitemShut
  {NoStop}%
\bibitem [{\citenamefont {Hussein}\ \emph {et~al.}(2007)\citenamefont
  {Hussein}, \citenamefont {Hulbert},\ and\ \citenamefont
  {Scott}}]{hussein2007dispersive}%
  \BibitemOpen
  \bibfield  {author} {\bibinfo {author} {\bibfnamefont {Mahmoud~I}\
  \bibnamefont {Hussein}}, \bibinfo {author} {\bibfnamefont {Gregory~M}\
  \bibnamefont {Hulbert}}, \ and\ \bibinfo {author} {\bibfnamefont {Richard~A}\
  \bibnamefont {Scott}},\ }\bibfield  {title} {\enquote {\bibinfo {title}
  {Dispersive elastodynamics of 1d banded materials and structures: design},}\
  }\href@noop {} {\bibfield  {journal} {\bibinfo  {journal} {Journal of Sound
  and Vibration}\ }\textbf {\bibinfo {volume} {307}},\ \bibinfo {pages}
  {865--893} (\bibinfo {year} {2007})}\BibitemShut {NoStop}%
\bibitem [{\citenamefont {Norris}(2008)}]{norris2008acoustic}%
  \BibitemOpen
  \bibfield  {author} {\bibinfo {author} {\bibfnamefont {A.N.}\ \bibnamefont
  {Norris}},\ }\bibfield  {title} {\enquote {\bibinfo {title} {Acoustic
  cloaking theory},}\ }\href@noop {} {\bibfield  {journal} {\bibinfo  {journal}
  {Proceedings of the Royal Society A: Mathematical, Physical and Engineering
  Science}\ }\textbf {\bibinfo {volume} {464}},\ \bibinfo {pages} {2411}
  (\bibinfo {year} {2008})}\BibitemShut {NoStop}%
\bibitem [{\citenamefont {Nayfeh}(1995)}]{nayfeh1995wave}%
  \BibitemOpen
  \bibfield  {author} {\bibinfo {author} {\bibfnamefont {Adnan}\ \bibnamefont
  {Nayfeh}},\ }\href@noop {} {\emph {\bibinfo {title} {Wave propagation in
  layered anisotropic media : with applications to composites}}}\ (\bibinfo
  {publisher} {Elsevier},\ \bibinfo {address} {Amsterdam New York},\ \bibinfo
  {year} {1995})\BibitemShut {NoStop}%
\bibitem [{\citenamefont {Willis}(2016)}]{willis2015negative}%
  \BibitemOpen
  \bibfield  {author} {\bibinfo {author} {\bibfnamefont {JR}~\bibnamefont
  {Willis}},\ }\bibfield  {title} {\enquote {\bibinfo {title} {Negative
  refraction in a laminate},}\ }\href@noop {} {\bibfield  {journal} {\bibinfo
  {journal} {Journal of the Mechanics and Physics of Solids}\ }\textbf
  {\bibinfo {volume} {97}},\ \bibinfo {pages} {10--18} (\bibinfo {year}
  {2016})}\BibitemShut {NoStop}%
\bibitem [{\citenamefont {Nemat-Nasser}(2015{\natexlab{b}})}]{NematNasser2015}%
  \BibitemOpen
  \bibfield  {author} {\bibinfo {author} {\bibfnamefont {Sia}\ \bibnamefont
  {Nemat-Nasser}},\ }\bibfield  {title} {\enquote {\bibinfo {title} {Anti-plane
  shear waves in periodic elastic composites: band structure and anomalous wave
  refraction},}\ }\href {\doibase 10.1098/rspa.2015.0152} {\bibfield  {journal}
  {\bibinfo  {journal} {Proceedings of the Royal Society A: Mathematical,
  Physical and Engineering Sciences}\ }\textbf {\bibinfo {volume} {471}},\
  \bibinfo {pages} {20150152} (\bibinfo {year}
  {2015}{\natexlab{b}})}\BibitemShut {NoStop}%
\bibitem [{\citenamefont {Norris}(1993)}]{norris1993waves}%
  \BibitemOpen
  \bibfield  {author} {\bibinfo {author} {\bibfnamefont {A.N.}\ \bibnamefont
  {Norris}},\ }\bibfield  {title} {\enquote {\bibinfo {title} {Waves in
  periodically layered media: a comparison of two theories},}\ }\href@noop {}
  {\bibfield  {journal} {\bibinfo  {journal} {SIAM Journal on Applied
  Mathematics}\ }\textbf {\bibinfo {volume} {53}},\ \bibinfo {pages} {1195}
  (\bibinfo {year} {1993})}\BibitemShut {NoStop}%
\bibitem [{\citenamefont {Norris}(1992)}]{norris1992dispersive}%
  \BibitemOpen
  \bibfield  {author} {\bibinfo {author} {\bibfnamefont {A.}~\bibnamefont
  {Norris}},\ }\bibfield  {title} {\enquote {\bibinfo {title} {Dispersive plane
  wave propagation in periodically layered anisotropic media},}\ }in\
  \href@noop {} {\emph {\bibinfo {booktitle} {Proceedings of the Royal Irish
  Academy. Section A: Mathematical and Physical Sciences}}},\ Vol.~\bibinfo
  {volume} {92}\ (\bibinfo {organization} {JSTOR},\ \bibinfo {year} {1992})\
  p.~\bibinfo {pages} {49}\BibitemShut {NoStop}%
\bibitem [{\citenamefont {Srivastava}(2016)}]{srivastava2016metamaterial}%
  \BibitemOpen
  \bibfield  {author} {\bibinfo {author} {\bibfnamefont {Ankit}\ \bibnamefont
  {Srivastava}},\ }\bibfield  {title} {\enquote {\bibinfo {title} {Metamaterial
  properties of periodic laminates},}\ }\href@noop {} {\bibfield  {journal}
  {\bibinfo  {journal} {Journal of the Mechanics and Physics of Solids}\
  }\textbf {\bibinfo {volume} {96}},\ \bibinfo {pages} {252--263} (\bibinfo
  {year} {2016})}\BibitemShut {NoStop}%
\bibitem [{\citenamefont {Amirkhizi}(2017)}]{amirkhizi2017homogenization}%
  \BibitemOpen
  \bibfield  {author} {\bibinfo {author} {\bibfnamefont {Alireza~V.}\
  \bibnamefont {Amirkhizi}},\ }\bibfield  {title} {\enquote {\bibinfo {title}
  {{Homogenization of layered media based on scattering response and field
  integration}},}\ }\href {\doibase 10.1016/J.MECHMAT.2017.06.008} {\bibfield
  {journal} {\bibinfo  {journal} {Mechanics of Materials}\ }\textbf {\bibinfo
  {volume} {114}},\ \bibinfo {pages} {76--87} (\bibinfo {year}
  {2017})}\BibitemShut {NoStop}%
\bibitem [{\citenamefont {Morini}\ \emph {et~al.}(2019)\citenamefont {Morini},
  \citenamefont {Eyzat},\ and\ \citenamefont {Gei}}]{Morini2019}%
  \BibitemOpen
  \bibfield  {author} {\bibinfo {author} {\bibfnamefont {Lorenzo}\ \bibnamefont
  {Morini}}, \bibinfo {author} {\bibfnamefont {Yoann}\ \bibnamefont {Eyzat}}, \
  and\ \bibinfo {author} {\bibfnamefont {Massimiliano}\ \bibnamefont {Gei}},\
  }\bibfield  {title} {\enquote {\bibinfo {title} {Negative refraction in
  quasicrystalline multilayered metamaterials},}\ }\href {\doibase
  10.1016/j.jmps.2018.10.016} {\bibfield  {journal} {\bibinfo  {journal}
  {Journal of the Mechanics and Physics of Solids}\ }\textbf {\bibinfo {volume}
  {124}},\ \bibinfo {pages} {282--298} (\bibinfo {year} {2019})}\BibitemShut
  {NoStop}%
\bibitem [{\citenamefont {Li}\ \emph {et~al.}(2018)\citenamefont {Li},
  \citenamefont {Slesarenko}, \citenamefont {Galich},\ and\ \citenamefont
  {Rudykh}}]{Li2018}%
  \BibitemOpen
  \bibfield  {author} {\bibinfo {author} {\bibfnamefont {Jian}\ \bibnamefont
  {Li}}, \bibinfo {author} {\bibfnamefont {Viacheslav}\ \bibnamefont
  {Slesarenko}}, \bibinfo {author} {\bibfnamefont {Pavel~I.}\ \bibnamefont
  {Galich}}, \ and\ \bibinfo {author} {\bibfnamefont {Stephan}\ \bibnamefont
  {Rudykh}},\ }\bibfield  {title} {\enquote {\bibinfo {title} {Oblique shear
  wave propagation in finitely deformed layered composites},}\ }\href {\doibase
  10.1016/j.mechrescom.2017.12.002} {\bibfield  {journal} {\bibinfo  {journal}
  {Mechanics Research Communications}\ }\textbf {\bibinfo {volume} {87}},\
  \bibinfo {pages} {21--28} (\bibinfo {year} {2018})}\BibitemShut {NoStop}%
\bibitem [{\citenamefont {Li}\ \emph {et~al.}(2019)\citenamefont {Li},
  \citenamefont {Slesarenko},\ and\ \citenamefont {Rudykh}}]{Li2019}%
  \BibitemOpen
  \bibfield  {author} {\bibinfo {author} {\bibfnamefont {Jian}\ \bibnamefont
  {Li}}, \bibinfo {author} {\bibfnamefont {Viacheslav}\ \bibnamefont
  {Slesarenko}}, \ and\ \bibinfo {author} {\bibfnamefont {Stephan}\
  \bibnamefont {Rudykh}},\ }\bibfield  {title} {\enquote {\bibinfo {title}
  {Microscopic instabilities and elastic wave propagation in finitely deformed
  laminates with compressible hyperelastic phases},}\ }\href {\doibase
  10.1016/j.euromechsol.2018.07.004} {\bibfield  {journal} {\bibinfo  {journal}
  {European Journal of Mechanics - A/Solids}\ }\textbf {\bibinfo {volume}
  {73}},\ \bibinfo {pages} {126--136} (\bibinfo {year} {2019})}\BibitemShut
  {NoStop}%
\bibitem [{\citenamefont {Srivastava}\ and\ \citenamefont
  {Willis}(2017)}]{srivastava2017evanescent}%
  \BibitemOpen
  \bibfield  {author} {\bibinfo {author} {\bibfnamefont {Ankit}\ \bibnamefont
  {Srivastava}}\ and\ \bibinfo {author} {\bibfnamefont {John~R}\ \bibnamefont
  {Willis}},\ }\bibfield  {title} {\enquote {\bibinfo {title} {Evanescent wave
  boundary layers in metamaterials and sidestepping them through a variational
  approach},}\ }in\ \href@noop {} {\emph {\bibinfo {booktitle} {Proc. R. Soc.
  A}}},\ Vol.\ \bibinfo {volume} {473}\ (\bibinfo {organization} {The Royal
  Society},\ \bibinfo {year} {2017})\ p.\ \bibinfo {pages}
  {20160765}\BibitemShut {NoStop}%
\bibitem [{\citenamefont {Sharma}\ and\ \citenamefont
  {Eremeyev}(2019)}]{sharma2019wave}%
  \BibitemOpen
  \bibfield  {author} {\bibinfo {author} {\bibfnamefont {Basant~Lal}\
  \bibnamefont {Sharma}}\ and\ \bibinfo {author} {\bibfnamefont {Victor~A}\
  \bibnamefont {Eremeyev}},\ }\bibfield  {title} {\enquote {\bibinfo {title}
  {Wave transmission across surface interfaces in lattice structures},}\
  }\href@noop {} {\bibfield  {journal} {\bibinfo  {journal} {International
  Journal of Engineering Science}\ }\textbf {\bibinfo {volume} {145}},\
  \bibinfo {pages} {103173} (\bibinfo {year} {2019})}\BibitemShut {NoStop}%
\bibitem [{\citenamefont {Mokhtari}\ \emph
  {et~al.}(2019{\natexlab{b}})\citenamefont {Mokhtari}, \citenamefont {Lu},\
  and\ \citenamefont {Srivastava}}]{mokhtari2019properties}%
  \BibitemOpen
  \bibfield  {author} {\bibinfo {author} {\bibfnamefont {Amir~Ashkan}\
  \bibnamefont {Mokhtari}}, \bibinfo {author} {\bibfnamefont {Yan}\
  \bibnamefont {Lu}}, \ and\ \bibinfo {author} {\bibfnamefont {Ankit}\
  \bibnamefont {Srivastava}},\ }\bibfield  {title} {\enquote {\bibinfo {title}
  {On the properties of phononic eigenvalue problems},}\ }\href@noop {}
  {\bibfield  {journal} {\bibinfo  {journal} {Journal of the Mechanics and
  Physics of Solids}\ }\textbf {\bibinfo {volume} {131}},\ \bibinfo {pages}
  {167--179} (\bibinfo {year} {2019}{\natexlab{b}})}\BibitemShut {NoStop}%
\bibitem [{\citenamefont {Makris}\ \emph {et~al.}(2008)\citenamefont {Makris},
  \citenamefont {El-Ganainy}, \citenamefont {Christodoulides},\ and\
  \citenamefont {Musslimani}}]{Makris2008}%
  \BibitemOpen
  \bibfield  {author} {\bibinfo {author} {\bibfnamefont {K.~G.}\ \bibnamefont
  {Makris}}, \bibinfo {author} {\bibfnamefont {R.}~\bibnamefont {El-Ganainy}},
  \bibinfo {author} {\bibfnamefont {D.~N.}\ \bibnamefont {Christodoulides}}, \
  and\ \bibinfo {author} {\bibfnamefont {Z.~H.}\ \bibnamefont {Musslimani}},\
  }\bibfield  {title} {\enquote {\bibinfo {title} {Beam dynamics
  {inPTSymmetric} optical lattices},}\ }\href {\doibase
  10.1103/physrevlett.100.103904} {\bibfield  {journal} {\bibinfo  {journal}
  {Physical Review Letters}\ }\textbf {\bibinfo {volume} {100}} (\bibinfo
  {year} {2008}),\ 10.1103/physrevlett.100.103904}\BibitemShut {NoStop}%
\bibitem [{\citenamefont {Kostenbauder}\ \emph {et~al.}(1997)\citenamefont
  {Kostenbauder}, \citenamefont {Sun},\ and\ \citenamefont
  {Siegman}}]{Kostenbauder1997}%
  \BibitemOpen
  \bibfield  {author} {\bibinfo {author} {\bibfnamefont {Adnah}\ \bibnamefont
  {Kostenbauder}}, \bibinfo {author} {\bibfnamefont {Yan}\ \bibnamefont {Sun}},
  \ and\ \bibinfo {author} {\bibfnamefont {A.~E.}\ \bibnamefont {Siegman}},\
  }\bibfield  {title} {\enquote {\bibinfo {title} {Eigenmode expansions using
  biorthogonal functions: complex-valued hermite{\textendash}gaussians},}\
  }\href {\doibase 10.1364/josaa.14.001780} {\bibfield  {journal} {\bibinfo
  {journal} {Journal of the Optical Society of America A}\ }\textbf {\bibinfo
  {volume} {14}},\ \bibinfo {pages} {1780} (\bibinfo {year}
  {1997})}\BibitemShut {NoStop}%
\bibitem [{\citenamefont {Lustig}\ \emph {et~al.}(2019)\citenamefont {Lustig},
  \citenamefont {Elbaz}, \citenamefont {Muhafra},\ and\ \citenamefont
  {Shmuel}}]{Gal2019}%
  \BibitemOpen
  \bibfield  {author} {\bibinfo {author} {\bibfnamefont {Ben}\ \bibnamefont
  {Lustig}}, \bibinfo {author} {\bibfnamefont {Guy}\ \bibnamefont {Elbaz}},
  \bibinfo {author} {\bibfnamefont {Alan}\ \bibnamefont {Muhafra}}, \ and\
  \bibinfo {author} {\bibfnamefont {Gal}\ \bibnamefont {Shmuel}},\ }\bibfield
  {title} {\enquote {\bibinfo {title} {Anomalous energy transport in laminates
  with exceptional points},}\ }\href {\doibase 10.1016/j.jmps.2019.103719}
  {\bibfield  {journal} {\bibinfo  {journal} {Journal of the Mechanics and
  Physics of Solids}\ }\textbf {\bibinfo {volume} {133}},\ \bibinfo {pages}
  {103719} (\bibinfo {year} {2019})}\BibitemShut {NoStop}%
\bibitem [{\citenamefont {Lu}\ and\ \citenamefont
  {Srivastava}(2018)}]{lu2018level}%
  \BibitemOpen
  \bibfield  {author} {\bibinfo {author} {\bibfnamefont {Y.}~\bibnamefont
  {Lu}}\ and\ \bibinfo {author} {\bibfnamefont {A.}~\bibnamefont
  {Srivastava}},\ }\bibfield  {title} {\enquote {\bibinfo {title} {{Level
  repulsion and band sorting in phononic crystals}},}\ }\href {\doibase
  10.1016/j.jmps.2017.10.021} {\bibfield  {journal} {\bibinfo  {journal}
  {Journal of the Mechanics and Physics of Solids}\ }\textbf {\bibinfo {volume}
  {111}} (\bibinfo {year} {2018}),\ 10.1016/j.jmps.2017.10.021}\BibitemShut
  {NoStop}%
\bibitem [{\citenamefont {Heiss}\ and\ \citenamefont
  {Sannino}(1990)}]{heiss1990avoided}%
  \BibitemOpen
  \bibfield  {author} {\bibinfo {author} {\bibfnamefont {WD}~\bibnamefont
  {Heiss}}\ and\ \bibinfo {author} {\bibfnamefont {AL}~\bibnamefont
  {Sannino}},\ }\bibfield  {title} {\enquote {\bibinfo {title} {Avoided level
  crossing and exceptional points},}\ }\href@noop {} {\bibfield  {journal}
  {\bibinfo  {journal} {Journal of Physics A: Mathematical and General}\
  }\textbf {\bibinfo {volume} {23}},\ \bibinfo {pages} {1167} (\bibinfo {year}
  {1990})}\BibitemShut {NoStop}%
\bibitem [{\citenamefont {Heiss}(2000)}]{heiss2000repulsion}%
  \BibitemOpen
  \bibfield  {author} {\bibinfo {author} {\bibfnamefont {WD}~\bibnamefont
  {Heiss}},\ }\bibfield  {title} {\enquote {\bibinfo {title} {Repulsion of
  resonance states and exceptional points},}\ }\href@noop {} {\bibfield
  {journal} {\bibinfo  {journal} {Physical Review E}\ }\textbf {\bibinfo
  {volume} {61}},\ \bibinfo {pages} {929} (\bibinfo {year} {2000})}\BibitemShut
  {NoStop}%
\bibitem [{\citenamefont {Ji}\ \emph {et~al.}(2019)\citenamefont {Ji},
  \citenamefont {Wei}, \citenamefont {Zhu}, \citenamefont {Wu},\ and\
  \citenamefont {Liu}}]{Ji2019}%
  \BibitemOpen
  \bibfield  {author} {\bibinfo {author} {\bibfnamefont {Wen-Qian}\
  \bibnamefont {Ji}}, \bibinfo {author} {\bibfnamefont {Qi}~\bibnamefont
  {Wei}}, \bibinfo {author} {\bibfnamefont {Xing-Feng}\ \bibnamefont {Zhu}},
  \bibinfo {author} {\bibfnamefont {Da-Jian}\ \bibnamefont {Wu}}, \ and\
  \bibinfo {author} {\bibfnamefont {Xiao-Jun}\ \bibnamefont {Liu}},\ }\bibfield
   {title} {\enquote {\bibinfo {title} {Extraordinary acoustic scattering in a
  periodic {PT}-symmetric zero-index metamaterials waveguide},}\ }\href
  {\doibase 10.1209/0295-5075/125/58002} {\bibfield  {journal} {\bibinfo
  {journal} {{EPL} (Europhysics Letters)}\ }\textbf {\bibinfo {volume} {125}},\
  \bibinfo {pages} {58002} (\bibinfo {year} {2019})}\BibitemShut {NoStop}%
\bibitem [{\citenamefont {El-Ganainy}\ \emph {et~al.}(2018)\citenamefont
  {El-Ganainy}, \citenamefont {Makris}, \citenamefont {Khajavikhan},
  \citenamefont {Musslimani}, \citenamefont {Rotter},\ and\ \citenamefont
  {Christodoulides}}]{ElGanainy2018}%
  \BibitemOpen
  \bibfield  {author} {\bibinfo {author} {\bibfnamefont {Ramy}\ \bibnamefont
  {El-Ganainy}}, \bibinfo {author} {\bibfnamefont {Konstantinos~G.}\
  \bibnamefont {Makris}}, \bibinfo {author} {\bibfnamefont {Mercedeh}\
  \bibnamefont {Khajavikhan}}, \bibinfo {author} {\bibfnamefont {Ziad~H.}\
  \bibnamefont {Musslimani}}, \bibinfo {author} {\bibfnamefont {Stefan}\
  \bibnamefont {Rotter}}, \ and\ \bibinfo {author} {\bibfnamefont
  {Demetrios~N.}\ \bibnamefont {Christodoulides}},\ }\bibfield  {title}
  {\enquote {\bibinfo {title} {Non-hermitian physics and {PT} symmetry},}\
  }\href {\doibase 10.1038/nphys4323} {\bibfield  {journal} {\bibinfo
  {journal} {Nature Physics}\ }\textbf {\bibinfo {volume} {14}},\ \bibinfo
  {pages} {11--19} (\bibinfo {year} {2018})}\BibitemShut {NoStop}%
\bibitem [{\citenamefont {Zhu}\ \emph {et~al.}(2014{\natexlab{b}})\citenamefont
  {Zhu}, \citenamefont {Ramezani}, \citenamefont {Shi}, \citenamefont {Zhu},\
  and\ \citenamefont {Zhang}}]{Zhu2014}%
  \BibitemOpen
  \bibfield  {author} {\bibinfo {author} {\bibfnamefont {Xuefeng}\ \bibnamefont
  {Zhu}}, \bibinfo {author} {\bibfnamefont {Hamidreza}\ \bibnamefont
  {Ramezani}}, \bibinfo {author} {\bibfnamefont {Chengzhi}\ \bibnamefont
  {Shi}}, \bibinfo {author} {\bibfnamefont {Jie}\ \bibnamefont {Zhu}}, \ and\
  \bibinfo {author} {\bibfnamefont {Xiang}\ \bibnamefont {Zhang}},\ }\bibfield
  {title} {\enquote {\bibinfo {title} {{PT}-symmetric acoustics},}\ }\href
  {\doibase 10.1103/physrevx.4.031042} {\bibfield  {journal} {\bibinfo
  {journal} {Physical Review X}\ }\textbf {\bibinfo {volume} {4}} (\bibinfo
  {year} {2014}{\natexlab{b}}),\ 10.1103/physrevx.4.031042}\BibitemShut
  {NoStop}%
\bibitem [{\citenamefont {Achenbach}(1984)}]{achenbach1984wave}%
  \BibitemOpen
  \bibfield  {author} {\bibinfo {author} {\bibfnamefont {Jan}\ \bibnamefont
  {Achenbach}},\ }\href@noop {} {\emph {\bibinfo {title} {Wave propagation in
  elastic solids}}}\ (\bibinfo  {publisher} {Elsevier},\ \bibinfo {year}
  {1984})\BibitemShut {NoStop}%
\bibitem [{\citenamefont {Rajabi}\ \emph {et~al.}(2020)\citenamefont {Rajabi},
  \citenamefont {Khodavirdi},\ and\ \citenamefont {Mojahed}}]{2002.00739}%
  \BibitemOpen
  \bibfield  {author} {\bibinfo {author} {\bibfnamefont {Majid}\ \bibnamefont
  {Rajabi}}, \bibinfo {author} {\bibfnamefont {Hossein}\ \bibnamefont
  {Khodavirdi}}, \ and\ \bibinfo {author} {\bibfnamefont {Alireza}\
  \bibnamefont {Mojahed}},\ }\href@noop {} {\enquote {\bibinfo {title}
  {Acoustic steering of active spherical carriers},}\ } (\bibinfo {year}
  {2020}),\ \Eprint {http://arxiv.org/abs/arXiv:2002.00739} {arXiv:2002.00739}
  \BibitemShut {NoStop}%
\bibitem [{\citenamefont {Achenbach}(2004)}]{Achenbach2004}%
  \BibitemOpen
  \bibfield  {author} {\bibinfo {author} {\bibfnamefont {J.~D.}\ \bibnamefont
  {Achenbach}},\ }\href {\doibase 10.1017/cbo9780511550485} {\emph {\bibinfo
  {title} {Reciprocity in Elastodynamics}}}\ (\bibinfo  {publisher} {Cambridge
  University Press},\ \bibinfo {year} {2004})\BibitemShut {NoStop}%
\bibitem [{\citenamefont {M\"{u}ller}\ and\ \citenamefont
  {Rotter}(2009)}]{Mller2009}%
  \BibitemOpen
  \bibfield  {author} {\bibinfo {author} {\bibfnamefont {Markus}\ \bibnamefont
  {M\"{u}ller}}\ and\ \bibinfo {author} {\bibfnamefont {Ingrid}\ \bibnamefont
  {Rotter}},\ }\bibfield  {title} {\enquote {\bibinfo {title} {Phase lapses in
  open quantum systems and the non-hermitian hamilton operator},}\ }\href
  {\doibase 10.1103/physreva.80.042705} {\bibfield  {journal} {\bibinfo
  {journal} {Physical Review A}\ }\textbf {\bibinfo {volume} {80}} (\bibinfo
  {year} {2009}),\ 10.1103/physreva.80.042705}\BibitemShut {NoStop}%
\bibitem [{\citenamefont {Persson}\ \emph {et~al.}(2000)\citenamefont
  {Persson}, \citenamefont {Rotter}, \citenamefont {St\"{o}ckmann},\ and\
  \citenamefont {Barth}}]{Persson2000}%
  \BibitemOpen
  \bibfield  {author} {\bibinfo {author} {\bibfnamefont {E.}~\bibnamefont
  {Persson}}, \bibinfo {author} {\bibfnamefont {I.}~\bibnamefont {Rotter}},
  \bibinfo {author} {\bibfnamefont {H.-J.}\ \bibnamefont {St\"{o}ckmann}}, \
  and\ \bibinfo {author} {\bibfnamefont {M.}~\bibnamefont {Barth}},\ }\bibfield
   {title} {\enquote {\bibinfo {title} {Observation of resonance trapping in an
  open microwave cavity},}\ }\href {\doibase 10.1103/physrevlett.85.2478}
  {\bibfield  {journal} {\bibinfo  {journal} {Physical Review Letters}\
  }\textbf {\bibinfo {volume} {85}},\ \bibinfo {pages} {2478--2481} (\bibinfo
  {year} {2000})}\BibitemShut {NoStop}%
\bibitem [{\citenamefont {Achilleos}\ \emph {et~al.}(2017)\citenamefont
  {Achilleos}, \citenamefont {Theocharis}, \citenamefont {Richoux},\ and\
  \citenamefont {Pagneux}}]{achilleos2017non}%
  \BibitemOpen
  \bibfield  {author} {\bibinfo {author} {\bibfnamefont {V}~\bibnamefont
  {Achilleos}}, \bibinfo {author} {\bibfnamefont {G}~\bibnamefont
  {Theocharis}}, \bibinfo {author} {\bibfnamefont {Olivier}\ \bibnamefont
  {Richoux}}, \ and\ \bibinfo {author} {\bibfnamefont {V}~\bibnamefont
  {Pagneux}},\ }\bibfield  {title} {\enquote {\bibinfo {title} {Non-hermitian
  acoustic metamaterials: Role of exceptional points in sound absorption},}\
  }\href@noop {} {\bibfield  {journal} {\bibinfo  {journal} {Physical Review
  B}\ }\textbf {\bibinfo {volume} {95}},\ \bibinfo {pages} {144303} (\bibinfo
  {year} {2017})}\BibitemShut {NoStop}%
\bibitem [{\citenamefont {Udwadia}(2012)}]{udwadia2012longitudinal}%
  \BibitemOpen
  \bibfield  {author} {\bibinfo {author} {\bibfnamefont {Firdaus~E}\
  \bibnamefont {Udwadia}},\ }\bibfield  {title} {\enquote {\bibinfo {title} {On
  the longitudinal vibrations of a bar with viscous boundaries:
  Super-stability, super-instability, and loss of damping},}\ }\href@noop {}
  {\bibfield  {journal} {\bibinfo  {journal} {International journal of
  engineering science}\ }\textbf {\bibinfo {volume} {50}},\ \bibinfo {pages}
  {79--100} (\bibinfo {year} {2012})}\BibitemShut {NoStop}%
\bibitem [{\citenamefont {Gao}\ \emph {et~al.}(2015)\citenamefont {Gao},
  \citenamefont {Estrecho}, \citenamefont {Bliokh}, \citenamefont {Liew},
  \citenamefont {Fraser}, \citenamefont {Brodbeck}, \citenamefont {Kamp},
  \citenamefont {Schneider}, \citenamefont {H{\"o}fling}, \citenamefont
  {Yamamoto} \emph {et~al.}}]{gao2015observation}%
  \BibitemOpen
  \bibfield  {author} {\bibinfo {author} {\bibfnamefont {Tiejun}\ \bibnamefont
  {Gao}}, \bibinfo {author} {\bibfnamefont {E}~\bibnamefont {Estrecho}},
  \bibinfo {author} {\bibfnamefont {KY}~\bibnamefont {Bliokh}}, \bibinfo
  {author} {\bibfnamefont {TCH}\ \bibnamefont {Liew}}, \bibinfo {author}
  {\bibfnamefont {MD}~\bibnamefont {Fraser}}, \bibinfo {author} {\bibfnamefont
  {Sebastian}\ \bibnamefont {Brodbeck}}, \bibinfo {author} {\bibfnamefont
  {Martin}\ \bibnamefont {Kamp}}, \bibinfo {author} {\bibfnamefont {Christian}\
  \bibnamefont {Schneider}}, \bibinfo {author} {\bibfnamefont {Sven}\
  \bibnamefont {H{\"o}fling}}, \bibinfo {author} {\bibfnamefont
  {Y}~\bibnamefont {Yamamoto}},  \emph {et~al.},\ }\bibfield  {title} {\enquote
  {\bibinfo {title} {Observation of non-hermitian degeneracies in a chaotic
  exciton-polariton billiard},}\ }\href@noop {} {\bibfield  {journal} {\bibinfo
   {journal} {Nature}\ }\textbf {\bibinfo {volume} {526}},\ \bibinfo {pages}
  {554--558} (\bibinfo {year} {2015})}\BibitemShut {NoStop}%
\bibitem [{\citenamefont {Frank}\ and\ \citenamefont {von
  Brentano}(1994)}]{frank1994classical}%
  \BibitemOpen
  \bibfield  {author} {\bibinfo {author} {\bibfnamefont {Winfried}\
  \bibnamefont {Frank}}\ and\ \bibinfo {author} {\bibfnamefont {Peter}\
  \bibnamefont {von Brentano}},\ }\bibfield  {title} {\enquote {\bibinfo
  {title} {Classical analogy to quantum mechanical level repulsion},}\
  }\href@noop {} {\bibfield  {journal} {\bibinfo  {journal} {American Journal
  of Physics}\ }\textbf {\bibinfo {volume} {62}},\ \bibinfo {pages} {706--709}
  (\bibinfo {year} {1994})}\BibitemShut {NoStop}%
\bibitem [{\citenamefont {Amirkhizi}\ and\ \citenamefont
  {Wang}(2018)}]{amirkhizi2018reduced}%
  \BibitemOpen
  \bibfield  {author} {\bibinfo {author} {\bibfnamefont {Alireza~V}\
  \bibnamefont {Amirkhizi}}\ and\ \bibinfo {author} {\bibfnamefont {Weidi}\
  \bibnamefont {Wang}},\ }\bibfield  {title} {\enquote {\bibinfo {title}
  {Reduced order derivation of the two-dimensional band structure of a
  mixed-mode resonator array},}\ }\href@noop {} {\bibfield  {journal} {\bibinfo
   {journal} {Journal of Applied Physics}\ }\textbf {\bibinfo {volume} {124}},\
  \bibinfo {pages} {245103} (\bibinfo {year} {2018})}\BibitemShut {NoStop}%
\bibitem [{\citenamefont {Miri}\ and\ \citenamefont
  {Al{\`u}}(2019)}]{miri2019exceptional}%
  \BibitemOpen
  \bibfield  {author} {\bibinfo {author} {\bibfnamefont {Mohammad-Ali}\
  \bibnamefont {Miri}}\ and\ \bibinfo {author} {\bibfnamefont {Andrea}\
  \bibnamefont {Al{\`u}}},\ }\bibfield  {title} {\enquote {\bibinfo {title}
  {Exceptional points in optics and photonics},}\ }\href@noop {} {\bibfield
  {journal} {\bibinfo  {journal} {Science}\ }\textbf {\bibinfo {volume}
  {363}},\ \bibinfo {pages} {eaar7709} (\bibinfo {year} {2019})}\BibitemShut
  {NoStop}%
\bibitem [{\citenamefont {Manconi}\ and\ \citenamefont
  {Mace}(2017)}]{manconi2017veering}%
  \BibitemOpen
  \bibfield  {author} {\bibinfo {author} {\bibfnamefont {Elisabetta}\
  \bibnamefont {Manconi}}\ and\ \bibinfo {author} {\bibfnamefont {Brian}\
  \bibnamefont {Mace}},\ }\bibfield  {title} {\enquote {\bibinfo {title}
  {Veering and strong coupling effects in structural dynamics},}\ }\href@noop
  {} {\bibfield  {journal} {\bibinfo  {journal} {Journal of Vibration and
  Acoustics}\ }\textbf {\bibinfo {volume} {139}} (\bibinfo {year}
  {2017})}\BibitemShut {NoStop}%
\bibitem [{\citenamefont {Manav}\ \emph {et~al.}(2014)\citenamefont {Manav},
  \citenamefont {Reynen}, \citenamefont {Sharma}, \citenamefont {Cretu},\ and\
  \citenamefont {Phani}}]{manav2014ultrasensitive}%
  \BibitemOpen
  \bibfield  {author} {\bibinfo {author} {\bibfnamefont {M}~\bibnamefont
  {Manav}}, \bibinfo {author} {\bibfnamefont {G}~\bibnamefont {Reynen}},
  \bibinfo {author} {\bibfnamefont {M}~\bibnamefont {Sharma}}, \bibinfo
  {author} {\bibfnamefont {E}~\bibnamefont {Cretu}}, \ and\ \bibinfo {author}
  {\bibfnamefont {AS}~\bibnamefont {Phani}},\ }\bibfield  {title} {\enquote
  {\bibinfo {title} {Ultrasensitive resonant mems transducers with tuneable
  coupling},}\ }\href@noop {} {\bibfield  {journal} {\bibinfo  {journal}
  {Journal of Micromechanics and Microengineering}\ }\textbf {\bibinfo {volume}
  {24}},\ \bibinfo {pages} {055005} (\bibinfo {year} {2014})}\BibitemShut
  {NoStop}%
\bibitem [{\citenamefont {Foreman}\ and\ \citenamefont
  {Vollmer}(2013)}]{foreman2013level}%
  \BibitemOpen
  \bibfield  {author} {\bibinfo {author} {\bibfnamefont {Matthew~R}\
  \bibnamefont {Foreman}}\ and\ \bibinfo {author} {\bibfnamefont {Frank}\
  \bibnamefont {Vollmer}},\ }\bibfield  {title} {\enquote {\bibinfo {title}
  {Level repulsion in hybrid photonic-plasmonic microresonators for enhanced
  biodetection},}\ }\href@noop {} {\bibfield  {journal} {\bibinfo  {journal}
  {Physical Review A}\ }\textbf {\bibinfo {volume} {88}},\ \bibinfo {pages}
  {023831} (\bibinfo {year} {2013})}\BibitemShut {NoStop}%
\bibitem [{\citenamefont {Dehrouyeh-Semnani}\ \emph {et~al.}(2016)\citenamefont
  {Dehrouyeh-Semnani}, \citenamefont {Mostafaei},\ and\ \citenamefont
  {Nikkhah-Bahrami}}]{dehrouyeh2016free}%
  \BibitemOpen
  \bibfield  {author} {\bibinfo {author} {\bibfnamefont {Amir~Mehdi}\
  \bibnamefont {Dehrouyeh-Semnani}}, \bibinfo {author} {\bibfnamefont {Hasan}\
  \bibnamefont {Mostafaei}}, \ and\ \bibinfo {author} {\bibfnamefont {Mansour}\
  \bibnamefont {Nikkhah-Bahrami}},\ }\bibfield  {title} {\enquote {\bibinfo
  {title} {Free flexural vibration of geometrically imperfect functionally
  graded microbeams},}\ }\href@noop {} {\bibfield  {journal} {\bibinfo
  {journal} {International Journal of Engineering Science}\ }\textbf {\bibinfo
  {volume} {105}},\ \bibinfo {pages} {56--79} (\bibinfo {year}
  {2016})}\BibitemShut {NoStop}%
\bibitem [{\citenamefont {Alcheikh}\ \emph {et~al.}(2019)\citenamefont
  {Alcheikh}, \citenamefont {Hajjaj},\ and\ \citenamefont
  {Younis}}]{alcheikh2019highly}%
  \BibitemOpen
  \bibfield  {author} {\bibinfo {author} {\bibfnamefont {Nouha}\ \bibnamefont
  {Alcheikh}}, \bibinfo {author} {\bibfnamefont {AZ}~\bibnamefont {Hajjaj}}, \
  and\ \bibinfo {author} {\bibfnamefont {Mohammad~I}\ \bibnamefont {Younis}},\
  }\bibfield  {title} {\enquote {\bibinfo {title} {Highly sensitive and
  wide-range resonant pressure sensor based on the veering phenomenon},}\
  }\href@noop {} {\bibfield  {journal} {\bibinfo  {journal} {Sensors and
  Actuators A: Physical}\ }\textbf {\bibinfo {volume} {300}},\ \bibinfo {pages}
  {111652} (\bibinfo {year} {2019})}\BibitemShut {NoStop}%
\bibitem [{\citenamefont {Hajjaj}\ \emph {et~al.}(2017)\citenamefont {Hajjaj},
  \citenamefont {Alcheikh},\ and\ \citenamefont {Younis}}]{hajjaj2017static}%
  \BibitemOpen
  \bibfield  {author} {\bibinfo {author} {\bibfnamefont {Amal~Z}\ \bibnamefont
  {Hajjaj}}, \bibinfo {author} {\bibfnamefont {Nouha}\ \bibnamefont
  {Alcheikh}}, \ and\ \bibinfo {author} {\bibfnamefont {Mohammad~I}\
  \bibnamefont {Younis}},\ }\bibfield  {title} {\enquote {\bibinfo {title} {The
  static and dynamic behavior of mems arch resonators near veering and the
  impact of initial shapes},}\ }\href@noop {} {\bibfield  {journal} {\bibinfo
  {journal} {International Journal of Non-Linear Mechanics}\ }\textbf {\bibinfo
  {volume} {95}},\ \bibinfo {pages} {277--286} (\bibinfo {year}
  {2017})}\BibitemShut {NoStop}%
\bibitem [{\citenamefont {Hodaei}\ \emph {et~al.}(2017)\citenamefont {Hodaei},
  \citenamefont {Hassan}, \citenamefont {Wittek}, \citenamefont
  {Garcia-Gracia}, \citenamefont {El-Ganainy}, \citenamefont
  {Christodoulides},\ and\ \citenamefont {Khajavikhan}}]{hodaei2017enhanced}%
  \BibitemOpen
  \bibfield  {author} {\bibinfo {author} {\bibfnamefont {Hossein}\ \bibnamefont
  {Hodaei}}, \bibinfo {author} {\bibfnamefont {Absar~U}\ \bibnamefont
  {Hassan}}, \bibinfo {author} {\bibfnamefont {Steffen}\ \bibnamefont
  {Wittek}}, \bibinfo {author} {\bibfnamefont {Hipolito}\ \bibnamefont
  {Garcia-Gracia}}, \bibinfo {author} {\bibfnamefont {Ramy}\ \bibnamefont
  {El-Ganainy}}, \bibinfo {author} {\bibfnamefont {Demetrios~N}\ \bibnamefont
  {Christodoulides}}, \ and\ \bibinfo {author} {\bibfnamefont {Mercedeh}\
  \bibnamefont {Khajavikhan}},\ }\bibfield  {title} {\enquote {\bibinfo {title}
  {Enhanced sensitivity at higher-order exceptional points},}\ }\href@noop {}
  {\bibfield  {journal} {\bibinfo  {journal} {Nature}\ }\textbf {\bibinfo
  {volume} {548}},\ \bibinfo {pages} {187--191} (\bibinfo {year}
  {2017})}\BibitemShut {NoStop}%
\bibitem [{\citenamefont {Heiss}(1999)}]{heiss1999phases}%
  \BibitemOpen
  \bibfield  {author} {\bibinfo {author} {\bibfnamefont {WD}~\bibnamefont
  {Heiss}},\ }\bibfield  {title} {\enquote {\bibinfo {title} {Phases of wave
  functions and level repulsion},}\ }\href@noop {} {\bibfield  {journal}
  {\bibinfo  {journal} {The European Physical Journal D-Atomic, Molecular,
  Optical and Plasma Physics}\ }\textbf {\bibinfo {volume} {7}},\ \bibinfo
  {pages} {1--4} (\bibinfo {year} {1999})}\BibitemShut {NoStop}%
\bibitem [{\citenamefont {Mailybaev}\ \emph {et~al.}(2005)\citenamefont
  {Mailybaev}, \citenamefont {Kirillov},\ and\ \citenamefont
  {Seyranian}}]{mailybaev2005geometric}%
  \BibitemOpen
  \bibfield  {author} {\bibinfo {author} {\bibfnamefont {Alexei~A}\
  \bibnamefont {Mailybaev}}, \bibinfo {author} {\bibfnamefont {Oleg~N}\
  \bibnamefont {Kirillov}}, \ and\ \bibinfo {author} {\bibfnamefont
  {Alexander~P}\ \bibnamefont {Seyranian}},\ }\bibfield  {title} {\enquote
  {\bibinfo {title} {Geometric phase around exceptional points},}\ }\href@noop
  {} {\bibfield  {journal} {\bibinfo  {journal} {Physical Review A}\ }\textbf
  {\bibinfo {volume} {72}},\ \bibinfo {pages} {014104} (\bibinfo {year}
  {2005})}\BibitemShut {NoStop}%
\bibitem [{\citenamefont {Thiruvenkatanathan}\ \emph
  {et~al.}(2009)\citenamefont {Thiruvenkatanathan}, \citenamefont {Yan},
  \citenamefont {Woodhouse}, \citenamefont {Aziz},\ and\ \citenamefont
  {Seshia}}]{thiruvenkatanathan2009effects}%
  \BibitemOpen
  \bibfield  {author} {\bibinfo {author} {\bibfnamefont {Pradyumna}\
  \bibnamefont {Thiruvenkatanathan}}, \bibinfo {author} {\bibfnamefont {Jize}\
  \bibnamefont {Yan}}, \bibinfo {author} {\bibfnamefont {Jim}\ \bibnamefont
  {Woodhouse}}, \bibinfo {author} {\bibfnamefont {A}~\bibnamefont {Aziz}}, \
  and\ \bibinfo {author} {\bibfnamefont {AA}~\bibnamefont {Seshia}},\
  }\bibfield  {title} {\enquote {\bibinfo {title} {Effects of mechanical and
  electrical coupling on the parametric sensitivity of mode localized
  sensors},}\ }in\ \href@noop {} {\emph {\bibinfo {booktitle} {2009 IEEE
  International Ultrasonics Symposium}}}\ (\bibinfo {organization} {IEEE},\
  \bibinfo {year} {2009})\ pp.\ \bibinfo {pages} {1183--1186}\BibitemShut
  {NoStop}%
\bibitem [{\citenamefont {Zhang}\ \emph {et~al.}(2017)\citenamefont {Zhang},
  \citenamefont {Zhong}, \citenamefont {Yang}, \citenamefont {Yuan},
  \citenamefont {Kang},\ and\ \citenamefont {Chang}}]{zhang2017algebraic}%
  \BibitemOpen
  \bibfield  {author} {\bibinfo {author} {\bibfnamefont {Hemin}\ \bibnamefont
  {Zhang}}, \bibinfo {author} {\bibfnamefont {Jiming}\ \bibnamefont {Zhong}},
  \bibinfo {author} {\bibfnamefont {Jing}\ \bibnamefont {Yang}}, \bibinfo
  {author} {\bibfnamefont {Weizheng}\ \bibnamefont {Yuan}}, \bibinfo {author}
  {\bibfnamefont {Hao}\ \bibnamefont {Kang}}, \ and\ \bibinfo {author}
  {\bibfnamefont {Honglong}\ \bibnamefont {Chang}},\ }\bibfield  {title}
  {\enquote {\bibinfo {title} {Algebraic summation of eigenstates as a novel
  output metric to extend the linear sensing range of mode-localized
  sensors},}\ }in\ \href@noop {} {\emph {\bibinfo {booktitle} {2017 IEEE
  SENSORS}}}\ (\bibinfo {organization} {IEEE},\ \bibinfo {year} {2017})\ pp.\
  \bibinfo {pages} {1--3}\BibitemShut {NoStop}%
\bibitem [{\citenamefont {Zhang}\ \emph
  {et~al.}(2018{\natexlab{a}})\citenamefont {Zhang}, \citenamefont {Kang},\
  and\ \citenamefont {Chang}}]{zhang2018suppression}%
  \BibitemOpen
  \bibfield  {author} {\bibinfo {author} {\bibfnamefont {Hemin}\ \bibnamefont
  {Zhang}}, \bibinfo {author} {\bibfnamefont {Hao}\ \bibnamefont {Kang}}, \
  and\ \bibinfo {author} {\bibfnamefont {Honglong}\ \bibnamefont {Chang}},\
  }\bibfield  {title} {\enquote {\bibinfo {title} {Suppression on nonlinearity
  of mode-localized sensors using algebraic summation of amplitude ratios as
  the output metric},}\ }\href@noop {} {\bibfield  {journal} {\bibinfo
  {journal} {IEEE Sensors Journal}\ }\textbf {\bibinfo {volume} {18}},\
  \bibinfo {pages} {7802--7809} (\bibinfo {year}
  {2018}{\natexlab{a}})}\BibitemShut {NoStop}%
\bibitem [{\citenamefont {Zhang}\ \emph
  {et~al.}(2018{\natexlab{b}})\citenamefont {Zhang}, \citenamefont {Yang},
  \citenamefont {Yuan},\ and\ \citenamefont {Chang}}]{zhang2018linear}%
  \BibitemOpen
  \bibfield  {author} {\bibinfo {author} {\bibfnamefont {Hemin}\ \bibnamefont
  {Zhang}}, \bibinfo {author} {\bibfnamefont {Jing}\ \bibnamefont {Yang}},
  \bibinfo {author} {\bibfnamefont {Weizheng}\ \bibnamefont {Yuan}}, \ and\
  \bibinfo {author} {\bibfnamefont {Honglong}\ \bibnamefont {Chang}},\
  }\bibfield  {title} {\enquote {\bibinfo {title} {Linear sensing for
  mode-localized sensors},}\ }\href@noop {} {\bibfield  {journal} {\bibinfo
  {journal} {Sensors and Actuators A: Physical}\ }\textbf {\bibinfo {volume}
  {277}},\ \bibinfo {pages} {35--42} (\bibinfo {year}
  {2018}{\natexlab{b}})}\BibitemShut {NoStop}%
\bibitem [{\citenamefont {Rabenimanana}\ \emph {et~al.}(2019)\citenamefont
  {Rabenimanana}, \citenamefont {Walter}, \citenamefont {Kacem}, \citenamefont
  {Le~Moal}, \citenamefont {Bourbon},\ and\ \citenamefont
  {Lardies}}]{rabenimanana2019mass}%
  \BibitemOpen
  \bibfield  {author} {\bibinfo {author} {\bibfnamefont {Toky}\ \bibnamefont
  {Rabenimanana}}, \bibinfo {author} {\bibfnamefont {Vincent}\ \bibnamefont
  {Walter}}, \bibinfo {author} {\bibfnamefont {Najib}\ \bibnamefont {Kacem}},
  \bibinfo {author} {\bibfnamefont {Patrice}\ \bibnamefont {Le~Moal}}, \bibinfo
  {author} {\bibfnamefont {Gilles}\ \bibnamefont {Bourbon}}, \ and\ \bibinfo
  {author} {\bibfnamefont {Joseph}\ \bibnamefont {Lardies}},\ }\bibfield
  {title} {\enquote {\bibinfo {title} {Mass sensor using mode localization in
  two weakly coupled mems cantilevers with different lengths: Design and
  experimental model validation},}\ }\href@noop {} {\bibfield  {journal}
  {\bibinfo  {journal} {Sensors and Actuators A: Physical}\ }\textbf {\bibinfo
  {volume} {295}},\ \bibinfo {pages} {643--652} (\bibinfo {year}
  {2019})}\BibitemShut {NoStop}%
\bibitem [{\citenamefont {Lang}\ \emph {et~al.}(2015)\citenamefont {Lang},
  \citenamefont {Liu},\ and\ \citenamefont {Monteiro}}]{lang2015dynamical}%
  \BibitemOpen
  \bibfield  {author} {\bibinfo {author} {\bibfnamefont {JE}~\bibnamefont
  {Lang}}, \bibinfo {author} {\bibfnamefont {Ren-Bao}\ \bibnamefont {Liu}}, \
  and\ \bibinfo {author} {\bibfnamefont {TS}~\bibnamefont {Monteiro}},\
  }\bibfield  {title} {\enquote {\bibinfo {title} {Dynamical-decoupling-based
  quantum sensing: Floquet spectroscopy},}\ }\href@noop {} {\bibfield
  {journal} {\bibinfo  {journal} {Physical Review X}\ }\textbf {\bibinfo
  {volume} {5}},\ \bibinfo {pages} {041016} (\bibinfo {year}
  {2015})}\BibitemShut {NoStop}%
\bibitem [{\citenamefont {Samutpraphoot}\ \emph {et~al.}(2020)\citenamefont
  {Samutpraphoot}, \citenamefont {Dordevic}, \citenamefont {Ocola},
  \citenamefont {Bernien}, \citenamefont {Senko}, \citenamefont {Vuleti{\'c}},\
  and\ \citenamefont {Lukin}}]{samutpraphoot2020strong}%
  \BibitemOpen
  \bibfield  {author} {\bibinfo {author} {\bibfnamefont {Polnop}\ \bibnamefont
  {Samutpraphoot}}, \bibinfo {author} {\bibfnamefont {Tamara}\ \bibnamefont
  {Dordevic}}, \bibinfo {author} {\bibfnamefont {Paloma~L}\ \bibnamefont
  {Ocola}}, \bibinfo {author} {\bibfnamefont {Hannes}\ \bibnamefont {Bernien}},
  \bibinfo {author} {\bibfnamefont {Crystal}\ \bibnamefont {Senko}}, \bibinfo
  {author} {\bibfnamefont {Vladan}\ \bibnamefont {Vuleti{\'c}}}, \ and\
  \bibinfo {author} {\bibfnamefont {Mikhail~D}\ \bibnamefont {Lukin}},\
  }\bibfield  {title} {\enquote {\bibinfo {title} {Strong coupling of two
  individually controlled atoms via a nanophotonic cavity},}\ }\href@noop {}
  {\bibfield  {journal} {\bibinfo  {journal} {Physical Review Letters}\
  }\textbf {\bibinfo {volume} {124}},\ \bibinfo {pages} {063602} (\bibinfo
  {year} {2020})}\BibitemShut {NoStop}%
\bibitem [{\citenamefont {Kundu}(2014)}]{kundu2014acoustic}%
  \BibitemOpen
  \bibfield  {author} {\bibinfo {author} {\bibfnamefont {Tribikram}\
  \bibnamefont {Kundu}},\ }\bibfield  {title} {\enquote {\bibinfo {title}
  {Acoustic source localization},}\ }\href@noop {} {\bibfield  {journal}
  {\bibinfo  {journal} {Ultrasonics}\ }\textbf {\bibinfo {volume} {54}},\
  \bibinfo {pages} {25--38} (\bibinfo {year} {2014})}\BibitemShut {NoStop}%
\bibitem [{\citenamefont {Haque}\ \emph {et~al.}(2017)\citenamefont {Haque},
  \citenamefont {Ghachi}, \citenamefont {Alnahhal}, \citenamefont {Aref},\ and\
  \citenamefont {Shim}}]{haque2017generalized}%
  \BibitemOpen
  \bibfield  {author} {\bibinfo {author} {\bibfnamefont {ABM~Tahidul}\
  \bibnamefont {Haque}}, \bibinfo {author} {\bibfnamefont {Ratiba~F}\
  \bibnamefont {Ghachi}}, \bibinfo {author} {\bibfnamefont {Wael~I}\
  \bibnamefont {Alnahhal}}, \bibinfo {author} {\bibfnamefont {Amjad}\
  \bibnamefont {Aref}}, \ and\ \bibinfo {author} {\bibfnamefont {Jongmin}\
  \bibnamefont {Shim}},\ }\bibfield  {title} {\enquote {\bibinfo {title}
  {Generalized spatial aliasing solution for the dispersion analysis of
  infinitely periodic multilayered composites using the finite element
  method},}\ }\href@noop {} {\bibfield  {journal} {\bibinfo  {journal} {Journal
  of Vibration and Acoustics}\ }\textbf {\bibinfo {volume} {139}},\ \bibinfo
  {pages} {051010} (\bibinfo {year} {2017})}\BibitemShut {NoStop}%
\bibitem [{\citenamefont {Wang}\ \emph {et~al.}(2019)\citenamefont {Wang},
  \citenamefont {Balogun},\ and\ \citenamefont {Achenbach}}]{achenbach2019}%
  \BibitemOpen
  \bibfield  {author} {\bibinfo {author} {\bibfnamefont {Chuanyong}\
  \bibnamefont {Wang}}, \bibinfo {author} {\bibfnamefont {Oluwaseyi}\
  \bibnamefont {Balogun}}, \ and\ \bibinfo {author} {\bibfnamefont {Jan~D.}\
  \bibnamefont {Achenbach}},\ }\bibfield  {title} {\enquote {\bibinfo {title}
  {Scattering of a rayleigh wave by a near surface crack which is normal to the
  free surface},}\ }\href {\doibase 10.1016/j.ijengsci.2019.103162} {\bibfield
  {journal} {\bibinfo  {journal} {International Journal of Engineering
  Science}\ }\textbf {\bibinfo {volume} {145}},\ \bibinfo {pages} {103162}
  (\bibinfo {year} {2019})}\BibitemShut {NoStop}%
\end{thebibliography}
%

\appendix
\section{Transfer Matrix Method (TMM)}
\label{appendix_A}
If in-plane waves are propagating in the laminate then the non-zero components of displacement is taken to be $u_1$, $u_2$ which has the functional form $u_1(x_1,x_2,t)$ and $u_2(x_1,x_2,t)$. This displacement field gives rise to stress fields $\sigma_{11}(x_1,x_2,t)$, $\sigma_{22}(x_1,x_2,t)$ and $\sigma_{12}(x_1,x_2,t)$. The stress component $\sigma_{11}$, $\sigma_{12}$ and displacement $u_1$, $u_2$ are continuous at the material interfaces. Across an interface between layers $i$ and $i+1$ at $x_1=x^i$:
\begin{equation}
\label{eContinuity}
\mathbf{v}^i|_{x_1=x^i}\equiv
\begin{pmatrix}
u_{1}(x^i,x_2,t) \\ 
u_2(x^i,x_2,t) \\
\sigma_{11}(x^i,x_2,t)\\
\sigma_{12}(x^i,x_2,t)
\end{pmatrix}^i=\mathbf{v}^{i+1}|_{x_1=x^i}
\end{equation}
Due to the periodicity of the laminate, the displacement and stress fields follow Bloch-periodicity conditions. By using the general solutions to the governing equations, the continuity of traction and displacement at the interfaces (\ref{eContinuity}), and the Bloch periodic boundary conditions, we can formulate a Transfer Matrix formulation ($x_2,\omega$ dependence suppressed):
\begin{equation}
\label{eTMM}
\mathbf{v}(h)=M(\omega,k_2)\mathbf{v}(0)=\lambda\mathbf{v}(0)
\end{equation}
where the eigenvalue $\lambda=e^{ik_1h}$. The eigenvalue solutions of the above come from the characteristic equation for $M$:
\begin{equation}
\label{eCharacteristic}
\lambda^4-a_3\lambda^3+a_2\lambda^2-a_1\lambda+a_0=0,
\end{equation}
where
\begin{eqnarray}
\label{app_param}
\nonumber a_3=\mathrm{tr}(M)\\
\nonumber a_2=\dfrac{1}{2}\left[\mathrm{tr}(M)^2-\mathrm{tr}(M^2)\right]\\
\nonumber a_1=\dfrac{1}{6}\left[\mathrm{tr}(M)^3-3\mathrm{tr}(M)\mathrm{tr}(M^2)+2\mathrm{tr}(M^3)\right]\\
a_0=\det(M).
\end{eqnarray}
These polynomial coefficients can be further simplified due to the symmetry of the layered composite
\begin{eqnarray}
\nonumber a_0=1\\
a_1=a_3=\mathrm{tr}(M).
\end{eqnarray}
The wave number solutions come from the following equation
\begin{equation}
\label{eInPlaneS}
\cos(k_1h)=\dfrac{1}{4}\left[a_3\pm\sqrt{a_3^2-4a_2+8}\right],
\end{equation}
so that if $k_1$ is a solution, then so are $\pm(k_1\pm 2n\pi/h)$ for all integer $n$ (details in \cite{haque2017generalized}).
\section{Betti-Rayleigh Reciprocity}

Betti-Rayleigh reciprocity \cite{Achenbach2004} can be used to relate two different admissible solutions of the elastodynamic problem in a region. Consider a volume $\Omega$ with boundary $\partial \Omega$. Two different sets of boundary conditions, body forces, and material distributions in $\Omega$ give rise to two different states of displacement ($\mathbf{u}$) and stress ($\boldsymbol{\sigma}$). For the current study, we are interested in the application of the Betti's theorem in the frequency domain. All quantities considered below are, therefore, understood to be in the frequency domain with the term $\exp(-i\omega t)$ uniformly suppressed. These sets are represented by superscripts $A,B$ and the governing equations for the two states are:
\begin{eqnarray}
\nonumber \displaystyle \sigma^A_{ij,j}+f^A_{i}+\omega^2\rho^Au^A_i=0\\
\displaystyle \sigma^B_{ij,j}+f^B_{i}+\omega^2\rho^Bu^B_i=0
\end{eqnarray}
Multiplying the first equation with $u^{B*}_i$, the second with $u^{A*}_i$, where $*$ denotes complex conjugation, and subtracting the two gives rise to the local version of the Betti's reciprocity theorem. This is converted to the global version by integrating over $\Omega$:
\begin{eqnarray}
\displaystyle \int_\Omega\left[\sigma_{ij,j}^Au_i^{B*}-\sigma_{ij,j}^Bu_i^{A*}\right]d\Omega+\int_\Omega\left[(f_i^A+\omega^2\rho^Au_i^A)u_i^{B*}-(f_i^B+\omega^2\rho^Bu_i^B)u_i^{A*}\right]d\Omega=0
\end{eqnarray}
which, in the absence of body forces, reduces to:
\begin{eqnarray}\label{eRec1}
\displaystyle \int_\Omega\left[\sigma_{ij,j}^Au_i^{B*}-\sigma_{ij,j}^Bu_i^{A*}\right]d\Omega+\omega^2\int_\Omega\left[\rho^Au_i^Au_i^{B*}-\rho^Bu_i^Bu_i^{A*}\right]d\Omega=0
\end{eqnarray}
The above will be the form of the reciprocity theorem that we use in this paper. The first integral will be further processed into a surface integral through the use of the Gauss theorem:
\begin{eqnarray}
\displaystyle \int_\Omega\left[\sigma_{ij,j}^Au_i^{B*}-\sigma_{ij,j}^Bu_i^{A*}\right]d\Omega=\int_{\partial\Omega}\left[\sigma_{ij}^Au_i^{B*}-\sigma_{ij}^Bu_i^{A*}\right]n_jdS-\int_\Omega\left[\sigma_{ij}^A\epsilon_{ij}^{B*}-\sigma_{ij}^B\epsilon_{ij}^{A*}\right]d\Omega
\end{eqnarray}
where $\epsilon_{ij}=u_{i,j}$. The above, when substituted into Eq. (\ref{eRec1}), results in the separation of surface and volume terms:
\begin{eqnarray}\label{eRec2}
\displaystyle \int_{\partial\Omega}\left[\sigma_{ij}^Au_i^{B*}-\sigma_{ij}^Bu_i^{A*}\right]n_jdS+\int_\Omega\left[\omega^2\rho^Au_i^Au_i^{B*}-\omega^2\rho^Bu_i^Bu_i^{A*}-\sigma_{ij}^A\epsilon_{ij}^{B*}+\sigma_{ij}^B\epsilon_{ij}^{A*}\right]d\Omega=0
\end{eqnarray}

For the unit cell considered in Fig. (\ref{fdomain1}) as $l\rightarrow 0$ this equation is simplified to:
\begin{eqnarray}\label{eRec4}
\displaystyle \int_{x_2=0^+}\left[t_{i}^Au_i^{B*}-t_{i}^Bu_i^{A*}\right]dS+\int_{x_2=0^-}\left[t_{i}^Au_i^{B*}-t_{i}^Bu_i^{A*}\right]dS=0
\end{eqnarray}
Since this is an in-plane problem, the non-zero displacement components are $u_1$, $u_2$ and the relevant stress quantities on planes parallel to the $x_1$ axis for the purpose of traction calculation are $\sigma_{21}$ and $\sigma_{22}$. The normal unit vector is positive on plane $x_2=0^+$ and is negative on $0^-$. Taking this into account the reciprocity relation becomes:
\begin{equation}\label{eRec41}
\int_{x_2=0^+}\left[\sigma_{21}^Au_1^{B*}+\sigma_{22}^Au_2^{B*}-\sigma_{21}^Bu_1^{A*}-\sigma_{22}^Bu_2^{A*}\right]dS+\int_{x_2=0^-}\left[-\sigma_{21}^Au_1^{B*}-\sigma_{22}^Au_2^{B*}+\sigma_{21}^Bu_1^{A*}+\sigma_{22}^Bu_2^{A*}\right]dS=0
\end{equation}
We now express the field variables in state $A$ in terms of the reflected and transmitted fields. Integration on $x_2=0^+,0^-$ collapse to $x_2=0$ with the appropriate representation of the transmitted and reflected fields respectively:
\begin{eqnarray}\label{eRec42}
\nonumber\int_{x_2=0}\left[\sigma_{21}^Tu_1^{B*}+\sigma_{22}^Tu_2^{B*}-\sigma_{21}^Bu_1^{T*}-\sigma_{22}^Bu_2^{T*}\right]dS\\
+\int_{x_2=0}\left[-(\sigma_{21}^I+\sigma_{21}^R)u_1^{B*}-(\sigma_{22}^I+\sigma_{22}^R)u_2^{B*}+\sigma_{21}^B(u_1^{I*}+u_1^{R*})+\sigma_{22}^B(u_2^{I*}+u_2^{R*})\right]dS=0
\end{eqnarray}
After some rearrangements, Eq. (\ref{eRec42}) can be written as
\begin{equation}\label{eBetti}
\int_{x_2=0} (\sigma_{2j}^T-\sigma_{2j}^R-\sigma_{2j}^{I})u_j^{B*}dx_1-\int_{x_2=0}(u_j^{T*}-u_j^{R*}-u_j^{I*})\sigma_{2j}^{B}dx_1=0,
\end{equation}
where summation convention is implied. The above is possible, in part, due to the fact that the free wave has continuous displacement and traction vectors across $x_2=0$. To solve the scattering problem, we first limit the infinite summation in Eq. (\ref{eScattered}) to a finite number of terms on the transmitted side ($N_t$) and on the reflected side $N_r$. Now we choose a set of $N_t+N_r$ virtual waves out of which $N_t$ waves are calculated by considering the virtual medium to be an infinite version of the layered composite and $N_r$ waves are calculated by considering the virtual medium to be an infinite version of the homogeneous medium. Substituting the the $ith$ virtual state with unit cell periodic displacement vector $\bar{\mathbf{u}}^{B,(i)}$ and stress tensor $\bar{\boldsymbol{\sigma}}^{B,(i)}$ into (\ref{eBetti}), we get the following:
\begin{eqnarray}\label{eq:Bettierecast}
\nonumber \int \sum_{m=1}^{N_t} T^{(m)}(\bar{\sigma}_{21}^{T,(m) }\bar{u}_{1}^{B,(i)*}+\bar{\sigma}_{22}^{T,(m)} \bar{u}_{2}^{B,(i)*}-\bar{u}_{1}^{T,(m)}\bar{\sigma}_{21}^{B,(i)*}-\bar{u}_{2}^{T,(m) }\bar{\sigma}_{22}^{B,(i)*})dx\\
\nonumber +\int \sum_{n=1}^{N_r} R^{(n)}(-\bar{\sigma}_{21}^{R,(n)} \bar{u}_1^{B,(i)*}-\bar{\sigma}_{22}^{R,(n)} \bar{u}_2^{B,(i)*}+\bar{u}_1^{R,(n)}\bar{\sigma}_{21}^{B,(i)*}+\bar{u}_2^{R,(n)}\bar{\sigma}_{22}^{B,(i)*})dx\\
=\int (\bar{\sigma}_{21}^{I}\bar{u}_1^{B,(i)*}+\bar{\sigma}_{22}^{I}\bar{u}_2^{B,(i)*}-\bar{u}_1^{I}\bar{\sigma}_{21}^{B,(i)*}-\bar{u}_2^{I}\bar{\sigma}_{22}^{B,(i)*})dx
\end{eqnarray}
By considering all the $N_t+N_r$ virtual sates in the above equation and after some algebraic manipulations, the following linear system of equation is obtained:
\begin{equation}
\mathbf{PS}=\mathbf{I}
\end{equation}
where $\mathbf{P}$ is a square matrix of size $(N_t+N_r)\times(N_t+N_r)$ and $\mathbf{S}=[T^{(1)}\quad T^{(2)}\quad...\quad T^{(N_t)}\quad R^{(1)}\quad R^{(2)}\quad ...\quad R^{(N_r)}]^T$ is the scattering coefficients vector. The $\mathbf{P}$ matrix consists of inner products of the periodic components of displacement and stress modeshapes.
\end{document}